\documentclass[preprint,3p]{elsarticle}
\journal{MSSP}
\usepackage{float}
\usepackage[utf8]{inputenc}
\usepackage{graphicx}           
\usepackage{amsmath}
\usepackage{amssymb}
\usepackage{xspace}
\usepackage{mathtools}
\usepackage{color}
\usepackage{hyperref}
\newcommand{\nc}{\newcommand}
\nc{\rnc}{\renewcommand}
\nc{\bs}{\boldsymbol}
\rnc{\matrix}[2]{\left[\!\!\begin{array}{#1}
	#2\end{array}\!\!\right]}
\rnc{\vector}[1]{\matrix{c}{#1}}
\nc{\mm}[1]{\boldsymbol{#1}}
\nc{\mms}[1]{\boldsymbol{#1}}
\nc{\real}[1]{\Re\left\{ #1 \right\}}
\nc{\imag}[1]{\Im\left\{ #1 \right\}}
\nc{\dd}{\mathrm{d}}
\nc{\ii}{\mathrm{i}}
\nc{\ee}{\mathrm{e}}
\nc{\inv}{^{-1}} 
\nc{\herm}{^{\mathrm H}}
\nc{\tra}{^{\mathrm T}}
\nc{\pinv}{^+}
\nc{\conj}[1]{ \overline{#1} }
\nc{\MM}{\mm M}
\nc{\BB}{\mm B}
\nc{\LL}{\mm L}
\nc{\ZZ}{\mm Z}
\nc{\eye}{\mm I}
\nc{\qq}{\mm q}
\nc{\dq}{\dot{\qq}}
\nc{\ddq}{\ddot{\qq}}
\nc{\fex}{\mm f}
\nc{\fexx}{\fex_{\mathrm{ex}}}
\nc{\fexxh}{\hat{\fex}_{\mathrm{ex}}}
\nc{\fcs}{c}
\nc{\fc}{\mm \fcs}
\nc{\fh}{\hat \fcs}
\nc{\fhnull}{\fh_0}
\nc{\xx}{\mm x}
\nc{\xr}{\tilde{\xx}}
\nc{\vv}{\mm\varphi}
\nc{\eex}{\mm e_{\fcs}}
\nc{\eextra}{\mm e\tra_{\fcs}}
\nc{\zA}{Z_{A}}
\nc{\zAst}{\zA^*}
\nc{\za}{Z_{\mathrm a}}
\nc{\ndof}{M}
\nc{\ndim}{N}
\nc{\ncon}{C}
\nc{\phiex}{\varphi_{\mathrm{ex}}}
\nc{\qtip}{q_{\mathrm{tip}}}
\nc{\qex}{q_{\mathrm{ex}}}
\nc{\qexh}{\hat q_{\mathrm{ex}}}
\nc{\mmod}{m_{\mathrm{mod}}}
\nc{\MOD}[1]{#1}
\nc{\COMMENT}[1]{\MOD{#1}}
\nc{\ie}{i.\,e.\xspace}
\nc{\eg}{e.\,g.\xspace}
\nc{\etc}{etc.\xspace}
\nc{\cf}{cf.\,}
\nc{\myquote}[1]{`#1'}
\nc{\etal}{et al.\xspace}
\nc{\fabstand}{\,}
\nc{\fp}{\fabstand.}
\nc{\fk}{\fabstand,}
\nc{\mrm}[1]{\mathrm{#1}} 

\nc{\exa}{\hat{\ddot w}_0}

\nc{\tab}[5][tbh]{\begin{table}[#1]\centering\caption{#4\label{tab:#5}}\begin{tabular}{#2}\hline #3 \\ \hline\end{tabular}\end{table}}
\nc{\fig}[4][tbh]{
\begin{figure}[#1]
\centering
\includegraphics[width=#4\textwidth]{figures/#2}
\caption{#3\label{fig:#2}}
\end{figure}}
\nc{\e}[2]{\begin{equation} #1 \label {eq:#2} \end{equation}}
\nc{\est}[1]{\begin{equation*} #1 \end{equation*}}
\nc{\ea}[1]{
\begin{eqnarray}
#1\end{eqnarray}}
\nc{\east}[1]{
\begin{eqnarray*}
#1 \end{eqnarray*}}
\nc{\fref}[1]{{Fig.~\ref{fig:#1}}}
\nc{\frefo}[1]{{\ref{fig:#1}}}
\nc{\frefs}[1]{{Figs.~\ref{fig:#1}}}
\nc{\tref}[1]{{Tab.~\ref{tab:#1}}}
\nc{\trefo}[1]{{\ref{tab:#1}}}
\nc{\trefs}[1]{{Tab.~\ref{tab:#1}}}
\nc{\eref}[1]{{Eq.~(\ref{eq:#1})}}
\nc{\erefo}[1]{(\ref{eq:#1})}
\nc{\erefs}[1]{{Eqs.~(\ref{eq:#1})}}
\nc{\sref}[1]{{Section~\ref{sec:#1}}}
\nc{\srefo}[1]{\ref{sec:#1}}
\nc{\srefs}[1]{{Sections~\ref{sec:#1}}}
\nc{\ssref}[1]{{Subsection~\ref{sec:#1}}}
\nc{\ssrefo}[1]{\ref{sec:#1}}
\nc{\ssrefs}[1]{{Subsections~\ref{sec:#1}}}
\nc{\aref}[1]{{{\ref{asec:#1}}}}
\nc{\arefo}[1]{{\ref{asec:#1}}}
\nc{\arefs}[1]{{{Appendices~\ref{asec:#1}}}}

\makeindex             

\begin{document}

\begin{frontmatter}
\title{
Experimental validation of a model for a self-adaptive beam-slider system
}
\author{Florian Müller$^1$}
\author{Maximilian W. Beck$^1$}
\author{Malte Krack$^1$}
\address{$^1$ University of Stuttgart, GERMANY}

\begin{abstract}
\MOD{
A system consisting of a doubly clamped beam with an attached body (slider) free to move along the beam has been studied recently by multiple research groups.
Under harmonic base excitation, the system has the capacity to passively adapt itself (by slowly changing the slider position) to yield either high or low vibrations.
The central contributions of this work are the refinement of the recently developed system model with regard to the finite stiffness of the beam's clamping, followed by a thorough validation of this model against experimental results.
With the intent to achieve repeatable and robust self-adaption, a new prototype of the system was designed, featuring, in particular, a continuously adjustable gap size and a concave inner contact geometry.
The initial beam model is updated based on the results of an Experimental Nonlinear Modal Analysis of the system (without slider).
By varying the excitation level and frequency in a wide range, all known types of behavior were reproduced in the experiment.
The simulation results of the updated model with slider are in excellent agreement with the measurements, both qualitatively (type of behavior) and quantitatively.
Minor deviations are attributed to the system's sensitivity to inevitable uncertainties, in particular with regard to the friction coefficient and the linear natural frequency.
It is thus concluded that the proposed model is well-suited for further analysis of its intriguing dynamics and for model-based optimization for technical applications such as energy harvesting.
}
\end{abstract}

\begin{keyword} 
Non-smooth dynamics, geometric nonlinearity, self-adaption, model updating, multiple scales
\end{keyword}

\end{frontmatter}
\section{Introduction}\label{sec:intro}
\begin{figure}[H]
	\begin{center}
		\includegraphics[width=0.85\textwidth]{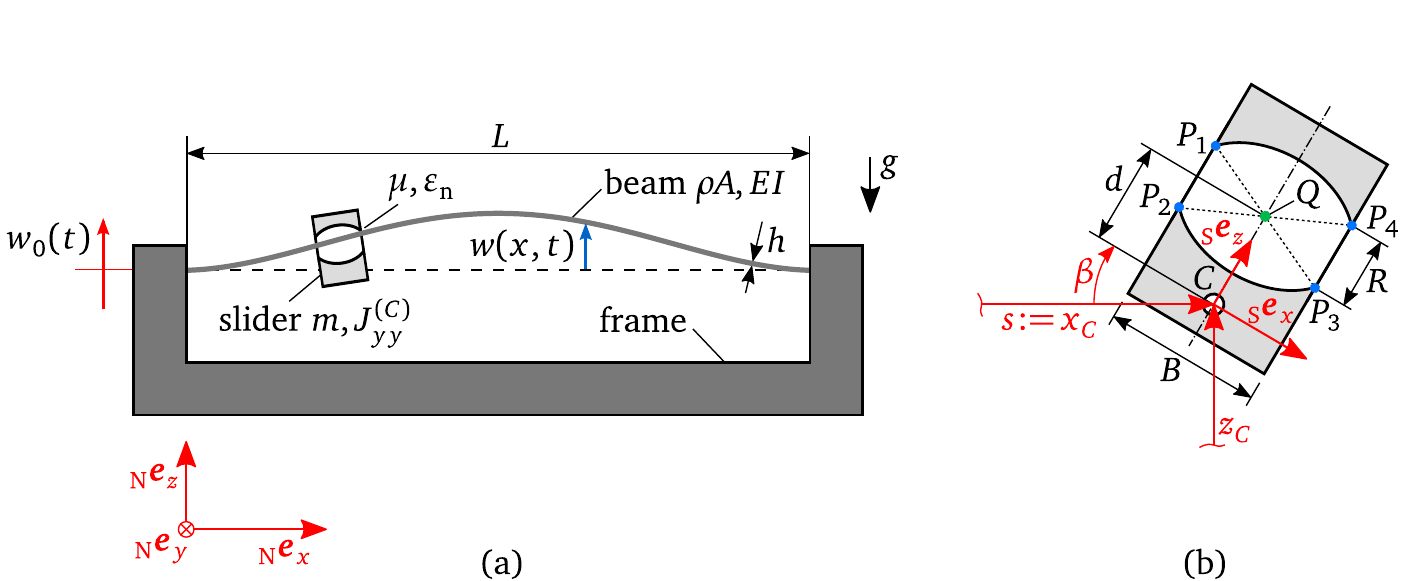}
		\caption{Schematic of self-adaptive system: (a) two-dimensional model of clamped-clamped beam with attached slider (b) slider detail}
		\label{fig:SAS_schematic}
	\end{center}
\end{figure}
%
Consider the beam-slider system schematically illustrated in \fref{SAS_schematic}a.
Under harmonic base excitation in the range around the beam's lowest-frequency bending mode, this system has shown self-adaptive behavior in experiments carried out independently by different research groups \cite{Miller.2013,Yu.2020,Shin.2020}.
The system was observed to adapt itself in such a way that it achieves and maintains high vibrations in a wide band of excitation frequencies.
After initially small vibrations, the slider moved to a certain position and the vibration level increased substantially.
Hereby a \emph{signature move} was observed~\cite{Miller.2013,Aboulfotoh.2017} (which is also shown later in~\fref{sign_m}) as described in the following.
The readers are strongly encouraged to watch the video \url{https://www.youtube.com/watch?v=qSy8ccbOgn8}.
While the system vibrates initially at small level, the slider moves towards the clamping.
At a certain point, the vibrations jump to a higher level and the slider turns back towards the beam's center.
This movement goes along with a further increase of the vibration level.
At a certain position, which can be different from the beam's center, the slider stops and large vibrations are maintained at steady state.
The above described passive re-adjustment to achieve and maintain large vibrations was achieved for a number of different material pairings and geometrical dimensions of the basic setup \cite{Miller.2013,Aboulfotoh.2017,Shin.2020}.
As this behavior occurs over a broad band of excitation frequencies in the vicinity of the lowest natural frequency, the described beam-slider system received considerable attention in the vibration energy harvesting community \MOD{\cite{Gregg.2014,Aboulfotoh.2017}.} 
\MOD{In addition to that, similar beam-slider systems with various boundary conditions of the beam were investigated in the context of energy harvesting \cite{Bukhari.2020,Lihua.2020,Lan.2021} or vibration suppression \cite{Yu.2022}. An overview on energy harvesting systems that are based on multistable mechanical systems is presented in \cite{Zhou.2022}.}

\subsubsection*{Vibration-induced sliding}
Although the setup of the beam-slider system is seemingly simple, and it has already been used in first energy harvesting applications, the self-adaptive behavior is not well-understood.
Apparently, the movement of the slider along the beam is caused by the beam's vibrations.
Consider first the case where the clearance between slider and beam is negligible compared to the beam's transverse vibration amplitude.
Then the slider's transverse displacement and rotation are constrained to those of the beam, $w(s,t)$ and $w^\prime(s,t)$, respectively, where $w^\prime = \partial w/\partial x$, $t$ is time and $s$ is the slider's axial position along the beam (\fref{SAS_schematic}a-b).
Assuming small displacements, slow movement of the slider and that the slider can be reasonably described as a point mass $m$, the equation governing its axial position reads \cite{Thomsen.1996}
\ea{
m\ddot s + m\left(\ddot w+\ddot w_0\right)w^{\prime} = 0\fp \label{eq:drivingForce}
}
Herein, overdot denotes derivative with respect to time, $\ddot w$ and $w^\prime$ are evaluated at $x=s$.
$\ddot w_0$ is the imposed base acceleration, $w$ is the elastic displacement. 
The second term on the left hand side of \eref{drivingForce} drives the slider along the beam and can be interpreted as transverse inertia force, $m\left(\ddot w+\ddot w_0\right)$, projected onto the rotated beam axis \cite{Miranda.1998}.
This driving force is kinematically nonlinear in the beam deformation.
The driving force in accordance with \eref{drivingForce} was used to explain the movement of the slider both in the case where the flexible host structure was a beam (\eg in \cite{Babitsky.1993,Khalily.1994}), and the case where it was a taut string (\eg in \cite{Thomsen.1996,Miranda.1998,Boudaoud.1999}). 
An immediate consequence of the above described theory is that equilibrium points are characterized by $w^\prime=0$ (zero slope), \eg, vibration anti-nodes \cite{Boudaoud.1999}. 
Even though the slider's mass was larger than that of the beam, and thus non-negligible, the vibrational deflection shape was clearly dominated by the lowest-frequency bending mode of the system without slider, regardless of the slider position, and thus had its anti-node approximately at the beam's center ($x=L/2$) \cite{Miller.2013,Muller.2020}.
The above theory can thus only explain the last part of the signature move, where the slider indeed moves towards the center.
In contrast, the movement away from the anti-node in the beginning of the signature move cannot be explained by the above theory.

\subsubsection*{Relevance of clearance and geometric nonlinearity}
The movement of the slider away from the anti-node was first reproduced via simulation in \cite{Krack.2017}.
In the beginning of the signature move, the vibrations are relatively small, such that the clearance between beam and slider cannot be neglected.
Hence, this clearance was explicitly taken into account in the model proposed in \cite{Krack.2017}.
The clearance gives rise to dynamical contact interactions at the four contact points indicated in \fref{SAS_schematic}b, which apparently leads to an intricate \emph{locomotion} process.
The locomotion process depends sensitively on the clearance.
If the clearance is too small, locomotion is hampered, as observed in both numerical investigations \cite{Krack.2017} and experiments \cite{Pillatsch.2013,Gregg.2014}.
Besides unilateral contact interactions, dry friction is known to be crucial: the slider would keep cycling back and forth along the beam in the frictionless case \cite{Krack.2017}.
\\
Due to the clamping present on both sides of the beam, bending deformation causes longitudinal stresses, which in turn increase the bending stiffness (string effect). 
As can be easily seen \eg in the video, the beam's transverse deformation is not small compared to its thickness.
This causes a considerable geometric hardening effect. 
Similar to the Duffing oscillator, the amplitude-frequency curve is strongly bent towards the right, so that the high-amplitude branch extends over a broad band of excitation frequencies.
This is viewed as cause for the broad-band efficacy intended for energy harvesting \cite{Yu.2020}.
Moreover, the geometric hardening nonlinearity leads to the co-existence of low- and high-level vibration states, and is required to explain the experimentally observed amplitude jumps \eg during the signature move \cite{Muller.2020} \MOD{(compare also \cite{Krack.2017}, where the beam model was linear and no amplitude jumps were observed).}

\subsubsection*{Limitations of the current knowledge and purpose of the present work}
The above described signature move is only one of several types of behavior of the beam-slider system.
The slider may also move monotonously towards the center (typically when it is initially placed near the clamping).
In this case, no amplitude jump occurs; high vibrations are also reached at steady state. \MOD{This type of behavior is referred as \emph{trivial adaption} in the following}, focussed on \eg in \citep{Yu.2019}, as opposed to the signature move. Both the signature move and the trivial adaption yield a steady state of high vibrations, and are in this sense forms of \emph{successful adaption}.
An intermittent behavior is also possible, where the slider undergoes a \emph{relaxation oscillation} along the beam (recurrent back and forth movement), with alternating high and low vibrations \cite{Gregg.2014,Muller.2020}.
This type of behavior is only rarely discussed in the literature.
Finally, the slider may show \emph{inactivity} (typically for low excitation), or move to a certain position leading to low steady-state vibrations \cite{Gregg.2014,Muller.2020}.
The latter type of behavior is sometimes referred to as \emph{self-damping} \cite{Babitsky.1993,Thomsen.1996,Miranda.1998}, as opposed to the successful self-adaption yielding high vibrations, which is also referred to as \emph{self-tuning} \cite{Boudaoud.1999,Miller.2013,Pillatsch.2013}.
\\
The conditions leading to a certain type of behavior are not understood yet.
In order to obtain this understanding, it is imperative to have a reliable mathematical model that is capable of accurately reproducing the experimental observations.
Remarkably, most proposed models of the beam-slider system ignore the clearance and the resulting contact interactions between beam and slider, and thus fail to qualitatively reproduce the locomotion of the slider.
The purpose of the present work is the quantitative experimental validation of a recently proposed model, which accounts both for the contact interactions and the beam's geometric nonlinearity \cite{Muller.2020}.
To this end, \MOD{a new prototype of the beam-slider system is designed}, as described in \sref{setup}, taking into account the experimental experience of \MOD{the authors} and other groups.
As it turns out, it is necessary to refine the model proposed in \cite{Muller.2020} to account for the finite compliance of the beam's clamping.
In \sref{new_beam_model}, the clamping stiffness is experimentally identified using the results of an Experimental Nonlinear Modal Analysis of the beam only (without slider).
The extension of the beam model by the slider and the modeling of the contact interactions between slider and beam is then described in \sref{model}.
Finally, simulation results are confronted with measurements in \sref{results} for a large range of operating conditions.
Concluding remarks are given in \sref{conclusions}.
\section{Experimental setup}\label{sec:setup}
For a thorough experimental validation of \MOD{the} simulation model, \MOD{a new prototype of the beam-slider system is developed}.
In the following, \MOD{the new design is described in detail, relevant parameters are specified and the instrumentation for the experimental investigations is explained}.
\subsection{Design of slider, beam and support structure}\label{sec:design}
Based on the scientific findings that \MOD{the authors} and other research groups made previously, \MOD{the following desired properties are specified}:
\begin{itemize}
\item[(1)] The beam should exhibit a pronounced geometric nonlinearity of hardening type \cite{Yu.2019,Yu.2019b,Muller.2020}.
\item[(2)] The clearance between beam and slider should be small (several percent of the beam's thickness) \cite{Miller.2013,Aboulfotoh.2017,Krack.2017,Muller.2020}. And it should be adjustable due to the known high sensitivity to this key parameter.
\item[(3)] The contact between slider and beam should be well-defined \cite{Krack.2017,Muller.2020}.
\item[(4)] The slider's center of mass should be away from the beam in vertical direction. This was found to increase robustness and speed of locomotion in preliminary simulations for this work.
\end{itemize}
From an engineering perspective, \MOD{three more desired properties are formulated}:
\begin{itemize}
\item[(5)] The beam should have high resistance against fatigue.
\item[(6)] The contact between slider and beam should be robust against wear.
\item[(7)] Uncertainties due to the assembly process should be avoided.
\end{itemize}
A photograph of the new prototype of the beam-slider system is shown in~\fref{exp_setup}a.
Both ends of the beam are clamped to a frame.
The free length of the beam is nominally straight with uniform rectangular cross section.
The frame is relatively stiff compared to the thin beam in order to enforce bending-stretching coupling which results in the desired geometric hardening  nonlinearity (1).
To be able to precisely adjust the gap size (2), the slider is made of two identical sliding parts and two identical couplers (\fref{exp_setup}b).
Before tightening the screws, the sliding parts can be moved closer together or away from each other in a continuous way.
The final clearance between slider and beam is adjusted by the aid of a feeler gauge.
To realize a well-defined contact geometry (3), the sliding part's side facing the beam is designed concave.
Consequently, the nominal contact interface consists of four lines (or four points in a planar model).
An additional threaded hole in the sliding parts enables to mount additional weights.
By adding weight to the lower side\MOD{, the center of mass is moved away from the beam} (4).
The beam is made of spring steel 1.8159 which provides a high fatigue strength (5).
To minimize contact wear (6), \MOD{sharp edges are avoided by designing} a rounded contact geometry of the sliding parts instead.
To avoid uncertainties during assembly (7), the clamping mechanism is as follows:
The inner screws 1 and 3 (as labelled in \fref{exp_setup}a) are M5x25 hexagon socket screws with the task to provide the normal force which clamps block, beam and frame together.
The outer screws 2 and 4 are fitting screws ISO 7379 with the task to position the beam and the block relative to the frame.
Screw 2 is forming a fit with the bores in block, beam and frame aligning the parts in~$x$-direction.
On the left side, the bore in the beam has a bigger diameter in order to avoid axial prestress in the beam, \ie screw~4 only aligns block and frame.
Also to minimize undesired prestresses, the screws are tightened in the order~$1\rightarrow2\rightarrow3\rightarrow4$.
In~\cite{Muller.2022,Abeloos.2022}\MOD{, the same frame and clamping mechanism were used} to analyze the softening-hardening behavior of a curved beam.
\\
\begin{figure}
	\begin{center}
		\includegraphics[width=0.55\textwidth]{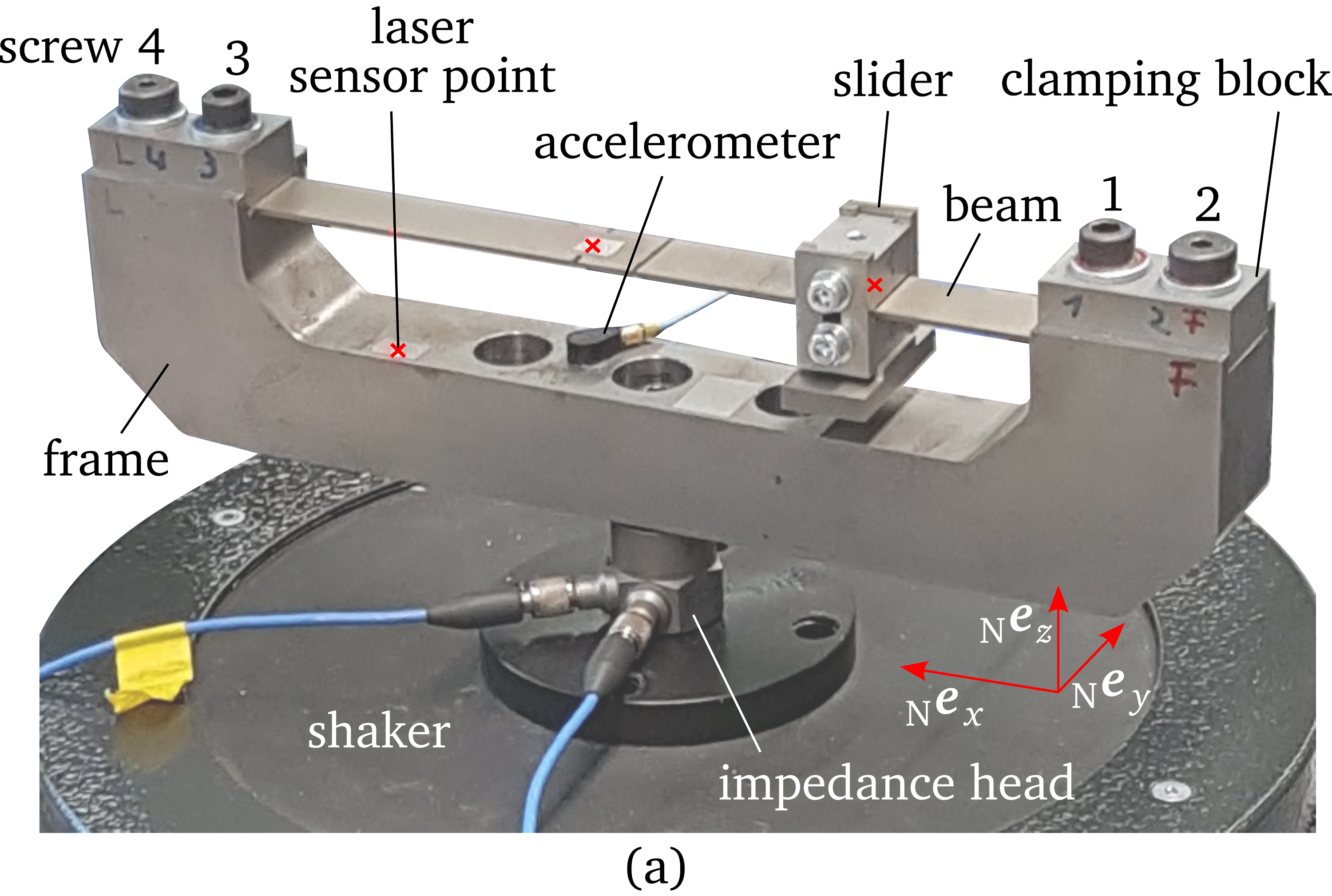}
		\includegraphics[width=0.37\textwidth]{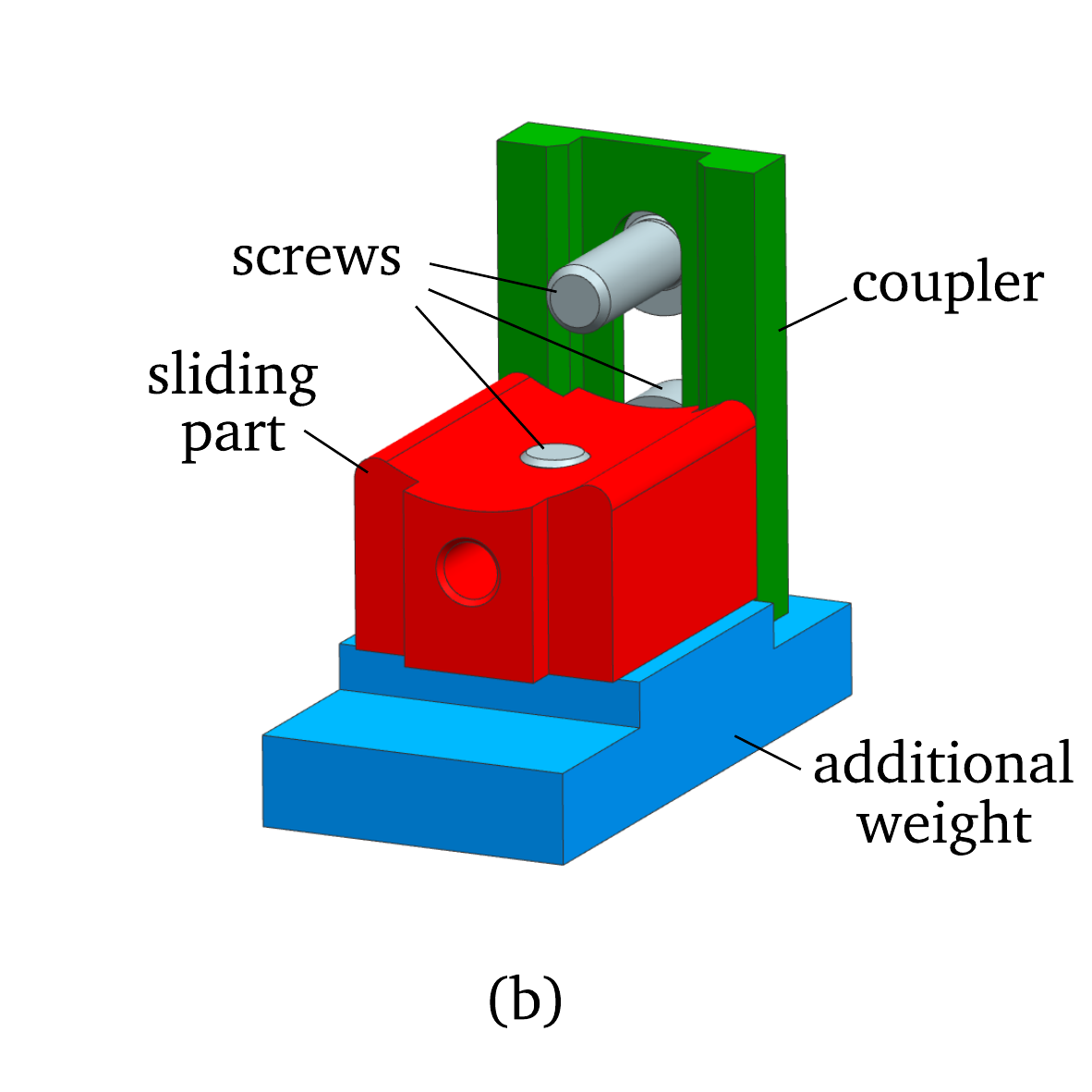}
		\caption{Experimental setup: (a) test specimen with instrumentation (b) CAD drawing of slider illustrating concave contact geometry, stepless adjustable gap mechanism and additional weight attachment (second sliding part and second coupler not shown for clarity)}
		\label{fig:exp_setup}
	\end{center}
\end{figure}
%
The beam and slider dimensions are close to the system investigated in~\cite{Aboulfotoh.2017}, where robust adaption was achieved in a broad frequency range.
The dimensions are large compared to, \eg, two of the setups in~\cite{Miller.2013}.
This makes it easier to adjust the slider's gap size relative to the beam's thickness.
Throughout the present work, the gap size was adjusted such that a clearance of 5\% of the beam's thickness is obtained.
This (relative) gap size was shown to enable successful adaption in~\citep{Muller.2020}.
The relevant parameters specifying the new prototype are collected in~\tref{parameters}.
\MOD{Any interested research group is welcome to} reproduce the experiments and/or use this as point of departure for future research.
Thus, \MOD{all technical drawings (including material specifications) and CAD files are provided} in a repository: \url{https://doi.org/10.18419/darus-2462} \cite{Muller.2022b}.
\begin{table}
	\caption{Parameters of the beam-slider setup}
	\label{tab:parameters}
	\begin{center}
		\begin{tabular}{llll}
			\hline
			Quantity & Symbol & Value & Unit\\
			\hline
			Free length (beam) & $L$ & 140 & $\mathrm{mm}$\\
			Thickness (beam) & $h$ & 1 & $\mathrm{mm}$\\
			Density (beam) & $\rho$ & 7683 & $\mathrm{kg/m^3}$\\
			Mass (beam, free length) & $\rho AL$ & 15.1 & $\mathrm{g}$\\
			Young's modulus (beam) & $E$ & 210 & $\mathrm{GPa}$\\
			Mass (slider) & $m$& 46.2 & $\mathrm{g}$\\
			Rotary inertia w.r.t. center of mass (slider) & $J_{yy}^{(C)}$ & 3.15 & $\mathrm{kg \ mm^2}$\\
			Distance between contact points (slider) & $B$ & 10 & $\mathrm{mm}$\\
			Distance from center of gravity (slider) & $d$ & 4.1 & $\mathrm{mm}$\\
			Relative gap size & $R/h$ & 1.05  & -- \\
			\hline
		\end{tabular}
	\end{center}
\end{table}
\subsection{Instrumentation of experiments with slider}\label{sec:instrumentation}
To realize base excitation, the frame was mounted onto an electrodynamic shaker (Brül \& Kjær type 4808, driven by amplifyer 2719).
Instead of screwing the frame directly to the shaker's armature, an impedance head (PCB model 288D01) was placed in between, see~\fref{exp_setup}a.
An accelerometer (PCB model 352C22) was attached to the frame's top side.
The accelerometer was used in preliminary tests only to verify the acceleration measured by the impedance head, which was in very  good agreement.
The beam's elastic deformation was measured by orienting the measurement head of a laser Doppler vibrometer (LDV) (Polytec OFV-552) toward the beam and the corresponding reference head toward the frame.
The slider is expected not to pass the beam's center.
To ensure that the optical access to the beam is not blocked by the slider, the laser was aligned to a position in the other half slightly away from the beam's center~($x=4L/7$).
The slider's horizontal ($x$) position was tracked by a triangulation sensor (Micro-Epsilon ILD1420-100).
For data acquisition and supply of the excitation signal (voltage signal operating the shaker via amplifier), all sensors and the amplifier are connected to a rapid control prototyping system (dSPACE microlab box).
This system was also used to control the amplitude of excitation to a prescribed acceleration value.
For this, the amplitude of the acceleration measured by the impedance head was evaluated online by synchronous demodulation (see~\cite{Abeloos.2022} for details).
This amplitude was controlled by adjusting the voltage amplitude of the harmonic excitation signal which was passed to the shaker's amplifier using a PI-controller (gains: $K_\mathrm{p}=0.03\mathrm{Vs}^2/\mathrm{m}, \ K_\mathrm{i}=0.13\mathrm{Vs}/\mathrm{m}$).
\MOD{The signal measured by the impedance head was used for this control task rather than the one of the accelerometer} due to the higher sensitivity of the impedance head.
\section{Refining and updating the model of the beam without slider}\label{sec:new_beam_model}
In \ssref{beam_model}, a refined mathematical model of the geometrically nonlinear beam is presented, which also accounts for the finite stiffness of the clamping.
The stiffness parameters and also the modal damping are updated as described in \ssref{updating}.
The updating procedure relies on the results of an Experimental Nonlinear Modal Analysis (ENMA).
The ENMA is described in \ssref{ENMA}.
\subsection{Geometrically nonlinear beam with finite-stiffness clamping: Equation of motion}\label{sec:beam_model}
In~\cite{Krack.2017,Muller.2020}, the beam's boundaries were idealized by assuming infinite stiffness of the clamping.
As will be shown in~\sref{updating}, this approximation yields large errors with regard to the linearized natural frequency and the extent of the hardening effect.
To account for the finite stiffness of the clamping in the model, \MOD{the boundary conditions are relaxed}, as illustrated in \fref{beam_model}.
\MOD{They are modeled} by hinges with rotational springs (stiffness~$k_\mathrm{r}$), where the right hinge is horizontally movable and constrained by a transversal spring (stiffness~$k_\mathrm{t}$).
\begin{figure}
	\begin{center}
		\includegraphics[width=0.61\textwidth]{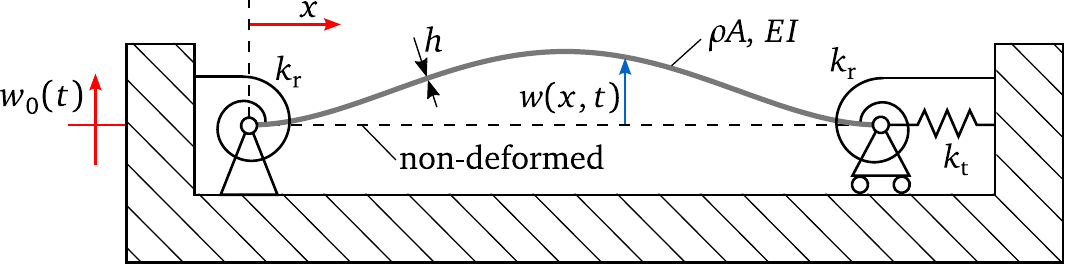}
		\caption{Model of beam with finite-stiffness clamping}
		\label{fig:beam_model}
	\end{center}
\end{figure}
\\
In the following, \MOD{the ordinary differential equation describing transverse elastic deformations of the beam is briefly derived}.
Using the conventional assumptions (small rotations, negligible longitudinal and rotational inertia, Hooke's law, Bernoulli assumptions, second-order kinematics), one can derive the partial differential equation,
\ea{
\rho A \ddot{w}+EI w^{\prime \prime \prime \prime} -H w^{\prime \prime}= -\rho A \ddot{w}_0\fk \label{eq:PDE}
}
for the case of an imposed base motion~$w_0(t)$.
Overdot and prime denote derivation with respect to time~$t$ and spatial coordinate~$x$, respectively.
$EI$ and $\rho A$ denote the bending stiffness and mass per unit length of the beam.
$H$ is the axial force which is determined as follows.
First, \MOD{the derivative of the horizontal deflection~$u(x,t)$ is considered}:
\ea{
u^\prime = \frac{H}{EA}-\frac{1}{2} \left( w^\prime\right )^2\fp\label{eq:bend_stretch}
}
Assuming that the horizontal motion of the right hinge is negligible compared to the beam's length, integrating over the beam's length and evaluating the resulting expression at the right end, where~$H=-k_\mathrm{t}u(L,t)$, yields
\ea{
H = \left(1+ \frac{EA}{L k_\mathrm{t}} \right)^{-1} \frac{EA}{2L} \int_0^L \left( w^\prime\right )^2 \dd x \fp
\label{eq:axial_force}
}
\MOD{Now, truncation to the $N$ lowest-frequency modes is applied,} using the ansatz
\ea{
w(x,t) \approx \sum_{n=1}^N \varphi_n(x) q_n(t) \fp \label{eq:separation}
}
Here, $\varphi_n(x)$ are the mass-normalized mode shapes of the underlying linearized system ($Hw^{\prime\prime}=0$, Euler-Bernoulli theory) and the boundary conditions
\ea{
w(0)=0 \fk \quad  EI w^{\prime \prime}(0) = k_\mathrm{r} w^\prime(0) \fk \quad w(L)=0 \fk \quad  EI w^{\prime \prime}(L) = -k_\mathrm{r} w^\prime(L) \fp \label{eq:BCs}
}
Substituting~\erefs{axial_force} and~\erefo{separation} into~\erefo{PDE}, and requiring that the residual is orthogonal with respect to the mode shapes, $\varphi_n$, yields a system of ordinary differential equations in terms of the sought modal coordinates, $q_n$,
\ea{
\ddot q_n + 2D_n \omega_n \dot q_n + \omega_n^2 q_n + \sum_{j=1}^N \sum_{k=1}^N \sum_{l=1}^N b_{j,k,l}^{(n)} q_j q_k q_l = \gamma_n \exa \cos(\Omega t) \quad \quad \mathrm{with} \quad n=1,2,\hdots,N\fp \label{eq:EOM_beam}
}
Herein, linear modal damping with damping ratios $D_n$ was also added, and the imposed base motion is assumed as harmonic, $w_0(t)=\exa \Omega^{-2} \cos(\Omega t)$.
$\omega_n$ is the natural (angular) frequency and $\gamma_n=\rho A \int_{0}^{L} \varphi_n \dd x$ the excitation coefficient of the~$n$-th linearized mode.
Note that equations \erefo{EOM_beam} are coupled by the nonlinear term describing bending-stretching coupling with the coefficients
\ea{
b_{j,k,l}^{(n)} = - \left(1+ \frac{EA}{L k_\mathrm{t}} \right)^{-1} \frac{EA}{2L} \int_0^L \varphi'_j \varphi'_k \mathrm{d}x \int_0^L \varphi''_l\varphi_n \dd x \fp \label{eq:coeffs}
}
Before \MOD{deriving the actual model updating procedure} in \ssref{updating}, \MOD{the Experimental Nonlinear Modal Analysis is explained} in \ssref{ENMA}, as this is used to gather the data basis for the model updating.
The focus will be placed on the fundamental bending mode.
For brevity, the associated quantities are simply expressed as $\omega := \omega_1$,~$\varphi(x):=\varphi_1(x)$,~$D:=D_1$ and $b := b_{1,1,1}^{(1)}$ in the following.
\subsection{Experimental Nonlinear Modal Analysis (ENMA) of the beam without slider}\label{sec:ENMA}
In this work, \MOD{the amplitude-dependent modal properties of the lowest frequency mode of the beam without slider are identified} from steady-state vibration data obtained under feedback-controlled phase-resonance testing.
The procedure is briefly described in the following; details can be found in \cite{Muller.2022}.
\\
The beam was clamped to the frame and mounted to a shaker (Brül \& Kjær type 4809, driven by amplifier 2718), similar to the setup described in~\sref{design} but without attaching the slider.
The target is to achieve phase resonance between the harmonic base excitation and the beam's transverse displacement relative to the base (response).
The excitation was measured by aligning a LDV (Polytec, OFV-505) to the left clamping block.
To measure the response, a second LDV (Polytec, OFV-552) with reference head was used. The measurement head of this second LDV was aligned to a point on the beam close to the clamping and the reference head to the left clamping block where also the excitation was measured.
To achieve phase resonance, a phase-locked loop was used with a PID controller (gains set as $K_\mathrm{p}=200 \ 1/\mathrm{s}$, $K_\mathrm{i}=100 \ 1/\mathrm{s}^2$, $K_\mathrm{d}=10$).
\\
Under the condition of phase resonance, the excitation frequency, $\Omega$, corresponds to the natural frequency, $\tilde{\omega}(a)=\Omega$, where $a$ is the modal amplitude and $\tilde{\omega}$ is the amplitude-dependent modal frequency.
By varying the excitation level, one can thus directly obtain the amplitude-dependent natural frequency.
In contrast, the amplitude-dependent modal damping ratio is more difficult to obtain, because the applied excitation force cannot be directly measured (to determine the power input to the structure that is in balance with the dissipated power).
\MOD{The method developed in \cite{Muller.2022} was used} to quantify the amplitude-dependent modal damping ratio using only response measurements.
More specifically, the following estimate was used:
\ea{
\tilde{D}(a) \approx \frac{\left| \hat{\bs w}^\mathrm{H} \bs b \exa \right|}{2 \hat{\bs w}^\mathrm{H} \hat{\bs w} \Omega^2} \fp \label{eq:damping}
}
Herein, $\hat{\bs w}$ is the complex-valued amplitude vector of the beam's transversal displacement relative to the base, $\bs b=[1,1,\hdots,1]^\top$ and $()^\mathrm{H}$ denotes the Hermitian (complex-conjugate transpose).
The estimation converges with the number of response sensors.
\MOD{A multi-point laser-Doppler vibrometer was used} to measure the response at five equidistant sensor points, $\hat{\bs w}=[\hat{w}(L/6), \hat{w}(2L/6),\hdots,\hat{w}(5L/6)]^\top$.
With this spatial resolution, the error made by the estimation in \eref{damping} was shown to be negligible for beams under different boundary conditions (pinned-pinned, clamped-clamped) \cite{Muller.2022}.
\\
The results of the ENMA are illustrated in \fref{ENMA}.
Here, the data is depicted for consecutive up-and-down stepping of the excitation level.
\MOD{The experiment was} started at low level and stepped upwards first (yielding result \#1), directly followed by stepping downwards (\#2).
The experiment was repeated after several minutes of rest (\#3 and \#4).
Each excitation level was held 20 seconds to reach steady state and record the last 300 periods.
The total test time of a consecutive up-and-down stepping was 920 seconds.
\begin{figure}
	\begin{center}
		\includegraphics[width=0.49\textwidth]{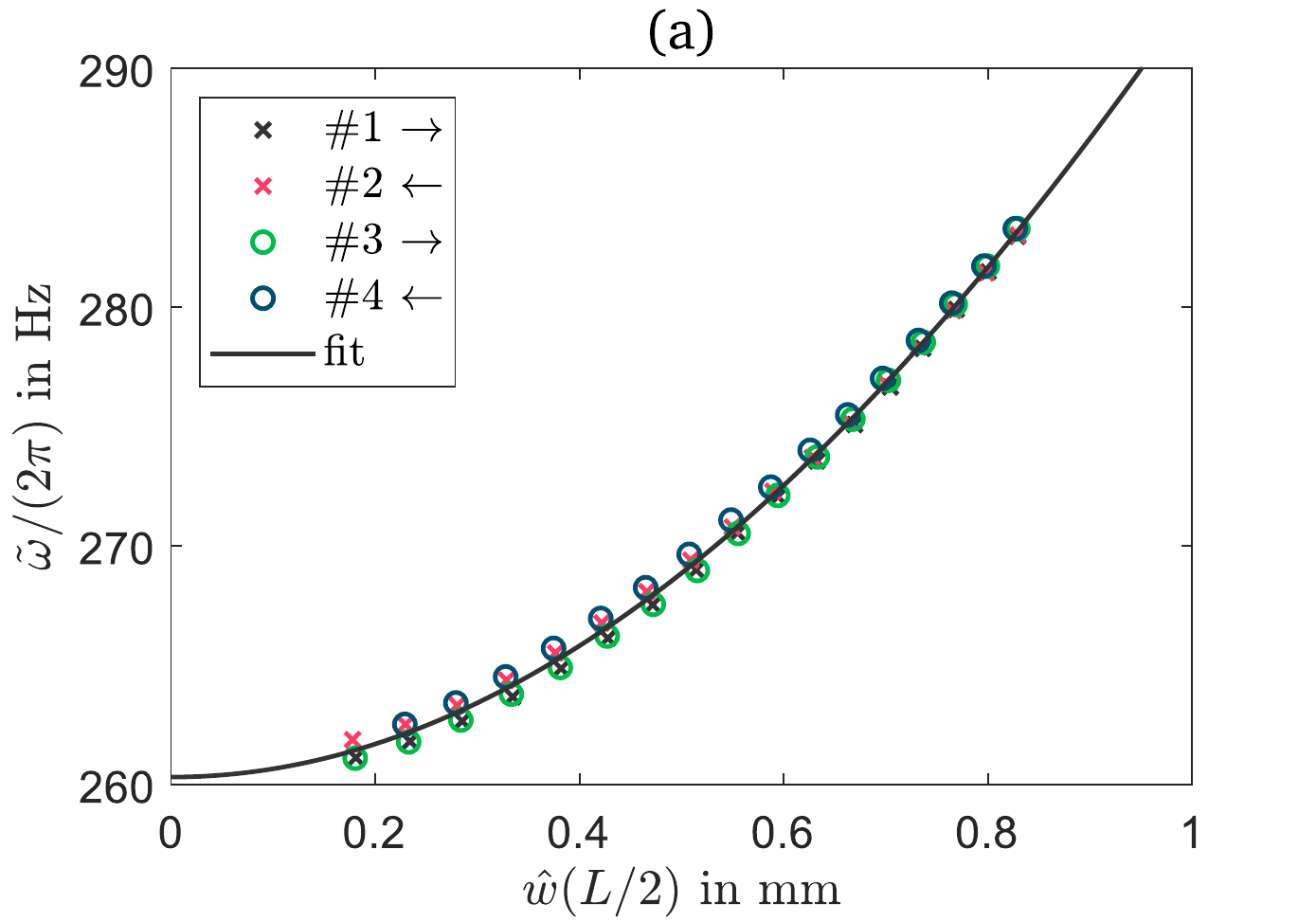}
		\includegraphics[width=0.49\textwidth]{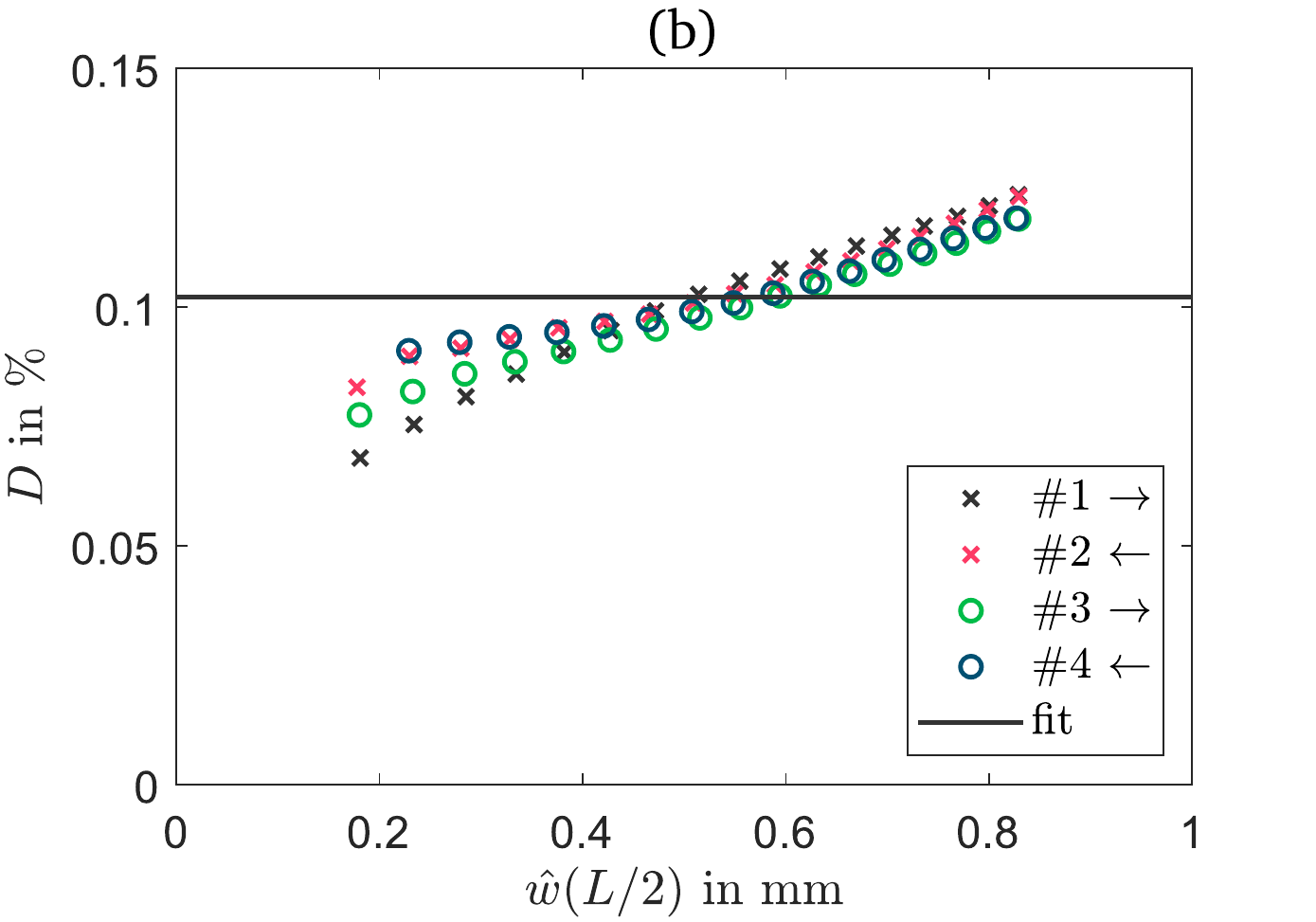}
		\includegraphics[width=0.49\textwidth]{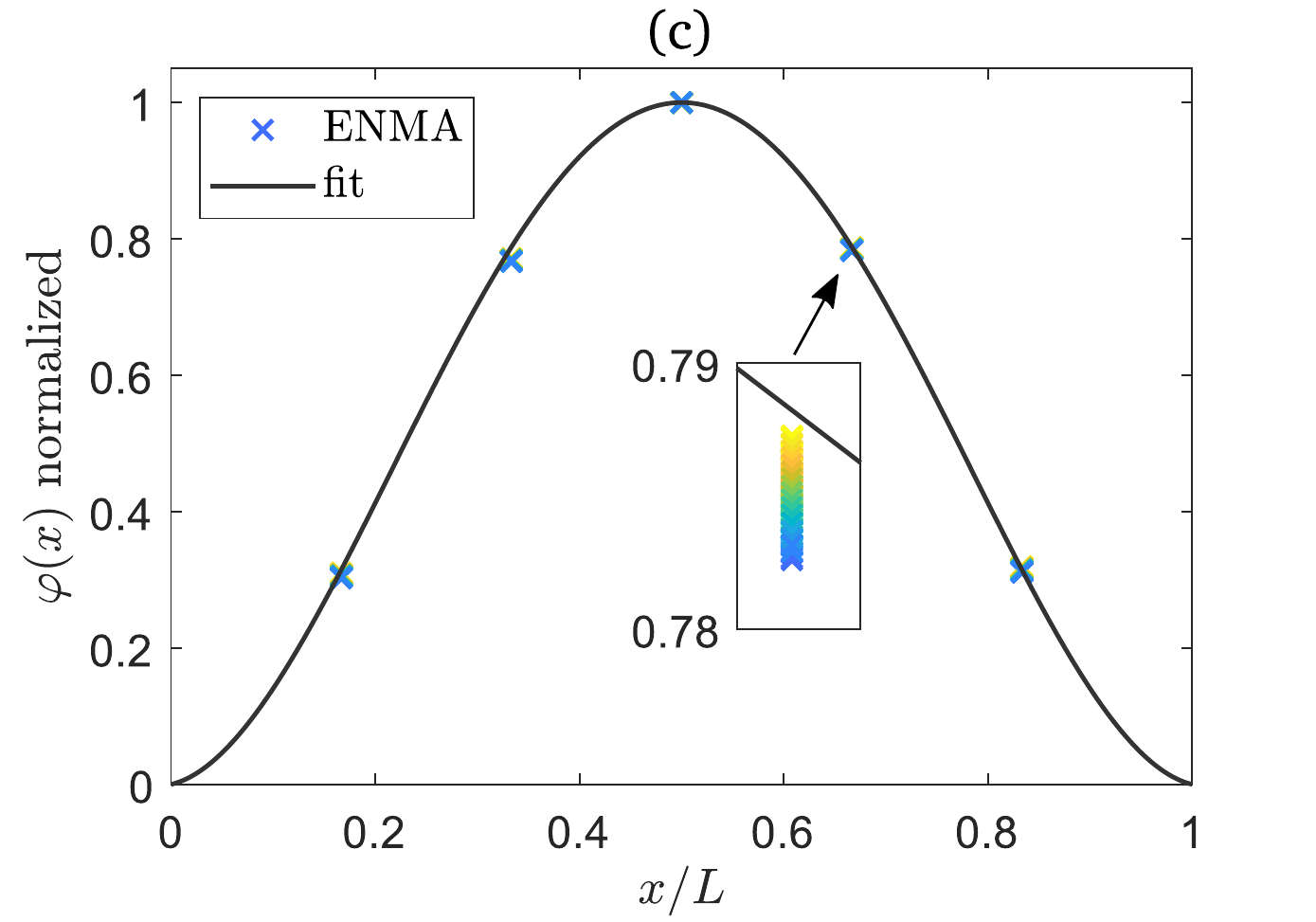}
		\caption{Results of ENMA and updated model: (a) modal frequency vs. amplitude, (b) modal damping ratio vs. amplitude,  (c) modal deflection shape, color of markers corresponds with amplitude level of experiment, blue: low level, yellow: high level}
		\label{fig:ENMA}
	\end{center}
\end{figure}
The curves \#1 and \#2 are slightly off at low level, which may be explained by an increase of temperature during the test.
The second run (\#3 and \#4) is in very good agreement with the first one.
The expected hardening behavior, which is dominant over the repetition variance, yields a frequency shift of more than 8\% in the considered amplitude range.
Damping is very light.
The slight increase with amplitude (less than 0.055\% of critical damping) is attributed to dry micro-slip friction in the bolted joints at the beam's clamping.
\\
The beam's modal deflection shape (normalized to unity) is shown in~\fref{ENMA}c.
Results of \#1--\#4 are plotted as markers, where the amplitude level is indicated by color.
Only a minor change of the deflection shape can be observed throughout the tests.
Consequently, the modal deflection shape is very similar to that of the linearized case at low amplitudes.
Moreover, the periodic vibration response is in good approximation harmonic:
The strongest higher harmonic of the measured velocity is ${<}5\%$ of the fundamental harmonic throughout all sensors and amplitude levels.
\subsection{Updating of the refined beam model}\label{sec:updating}
The dominance of the (single) linearized modal deflection shape and the fundamental temporal harmonic, observed in the ENMA, allows one to pursue a rather simple model updating procedure.
In particular, truncating \eref{EOM_beam} to a single linearized mode, $N=1$, and applying single-term Harmonic Balance yields a closed-form expression for the amplitude-dependent natural frequency (see \eg~\cite{nayf1979}):
\ea{
\tilde{\omega}^2(\tilde{a})=\omega^2+\frac{3}{4}b^*\tilde{a}^2\fk  \label{eq:om_SHB}
}
with
\ea{
\quad b^*:=\frac{bh^2}{ \left(\varphi(L/2) \right)^{2}} \fk \quad \tilde{a}:=\frac{\hat{w}(L/2)}{h}\fp \label{eq:normalization}
}
From the ENMA, $\tilde{\omega}^2(\tilde{a})$ is known, and this allows \MOD{one} to determine the linear natural frequency $\omega$, and nonlinear coefficient $b^*$ via \eref{om_SHB} using a least-squares fit:
\ea{
\begin{bmatrix}\omega^2 \\ b^*\end{bmatrix} \approx
\begin{bmatrix}1 & \frac{3}{4} \tilde{a}_1^2 \\ 1 & \frac{3}{4} \tilde{a}_2^2 \\ \vdots & \vdots \\ 1 & \frac{3}{4} \tilde{a}_K^2 \end{bmatrix}\pinv
\begin{bmatrix}\tilde{\omega}_1^2 \\ \tilde{\omega}_2^2 \\ \vdots \\ \tilde{\omega}_K^2\end{bmatrix} \fp
\label{eq:fit_om_b}
}
Here~$K$ is the total number of phase-resonant points measured during the ENMA and~$()^+$ denotes the Moore-Penrose pseudo inverse.
Evaluating~\eref{om_SHB} with fitted~$\omega$ and~$b^*$ yields excellent agreement with the measured data (\cf~\fref{ENMA}a), which further supports the validity of the single-mode-single-harmonic truncation.
\\
Next, \MOD{it is assumed} that the model uncertainty resides in the clamping stiffness parameters, $k_\mathrm{r}$, $k_\mathrm{t}$, whereas the beam's geometric and material properties correspond to the nominal ones as specified in \tref{parameters}.
The underlying linear model is not affected by the transversal elasticity~$k_\mathrm{t}$.
Hence, the rotational spring~$k_\mathrm{r}$ can be simply determined by solving
\ea{
k_\mathrm{r} = \mathrm{argmin} \left( \left|  \omega - \omega_\mathrm{mod}(k_\mathrm{r}) \right| \right)=\MOD{125.0}\frac{\mathrm{Nm}}{\mathrm{rad}} \fk \label{eq:k_r}
}
where $\omega$ is the fitted value and $\omega_\mathrm{mod}(k_\mathrm{r})$ is the lowest natural frequency of the linear Euler-Bernoulli beam in accordance with the boundary conditions specified in~\eref{BCs}.
Subsequently, the stiffness of the transversal spring can be determined from~\eref{coeffs} yielding
\ea{
k_\mathrm{t} = \frac{EA}{L \left( {b^*_\infty}/{b^*}-1 \right)}=1.93\cdot10^7\frac{\mathrm{N}}{\mathrm{m}} \fp \label{eq:k_t}
}
Here~$b^*_\infty$ is the value with neglected horizontal elasticity of the frame (obtained from~\eref{coeffs} with~$k_\mathrm{t} \to \infty$ and normalization as in~\eref{normalization}) and $b^*$ is the fitted value.
Compared to the beam model with ideal clamping ($k_\mathrm{r}, \ k_\mathrm{t} \to \infty$), the obtained linear frequency~$\omega$ and nonlinear stiffness coefficient~$b^*$ are 5.06\% and 52.17\% lower respectively.
This shows that a model with ideal clamping is not suitable to accurately describe the dynamic behavior of the beam without slider.
\\
After having found~$k_\mathrm{r}$ and~$k_\mathrm{t}$, the model is fully determined, except for the damping.
The measured damping is generally weak and has only moderate amplitude dependency.
It is expected that this beam-only damping plays a minor role once the slider is attached, because then unilateral and frictional contact interactions are expected to provide substantial dissipation.
For this reason, \MOD{the beam-only damping is simply modeled} as linear modal damping, \MOD{using} the mean value of the damping ratio from the ENMA, $D_1=0.102\%$.
With the same argument, \MOD{this damping ratio is used} also for the other modes; \ie, $D_n=D_1$ for $n=2,\ldots,N$.
\section{Modeling the contact interactions with the slider and simulation of the coupled beam-slider system}\label{sec:model}
Except for the refinement of the beam model with regard to the finite-stiffness clamping, modeling and simulation approach are largely adopted from \MOD{the authors'} previous work \cite{Krack.2017,Muller.2020}.
For self-consistency of the present paper, a brief recap is given in this section along with the specifics of the model and the simulations.
\subsection{Slider model, contact model, kinematics and numerical simulation\label{sec:EoM}}
The system is idealized as a planar one and the slider is modelled as rigid body with horizontal, vertical and rotational degrees of freedom (\fref{SAS_schematic}).
Gravity is taken into account as indicated in \fref{SAS_schematic} in the slider model\footnote{It was found that the effect of gravity on the beam (due to its own weight) negligible.}.
The slider's inner geometry is such that contact may occur only at the four points indicated in \fref{SAS_schematic}b.
Considering the contact kinematics, \MOD{the approximation~$x_i^*\approx x_{P_i}$ is newly introduced}. Here $x_{P_i}$ is the horizontal coordinate of the considered contact point~$P_i$ of the slider and $x_i^*$ the coordinate of the point one obtains when projecting~$P_i$ onto the beam's center line. As a consequence of small rotations, this approximation has negligible effect on the simulation results, which was ensured by a prestudy. The approximation makes the numerical solution for~$x_i^*$ obsolete, which reduces the overall computation cost by about 50\%.
Unilateral and dry frictional contact interactions between beam and slider are modeled by the Signorini and Coulomb laws combined with Newton's impact law.
These are set-valued contact laws, which are associated with velocity jumps, and thus standard time integration algorithms for continuous ordinary differential equation systems cannot be applied.
Instead, the equation of motion is transformed to a measure differential inclusion, which is solved numerically by Moreau's time stepping scheme (see \eg~\cite{Studer.2009}).
The dynamic contact problem is resolved using an augmented Lagrangian approach.
In accordance with preliminary convergence studies, the beam model is truncated to $N=5$ modes\footnote{Besides the modal convergence of the depicted results it was also verified that the retained set of beam-only modes can properly represent the mode shape of the beam with rigidly attached mass.} and a fixed time step of $\Delta t =2{\cdot}10^{-5}\mrm{s}$ is used in all simulations.
This time step size is sufficiently small to resolve a vibration period corresponding to the highest linear frequency ($\omega_5$) with ten time levels.
\subsection{Contact parameters}\label{sec:sim_prms}
The coefficient of friction was determined experimentally as follows.
The frame was inclined slowly and the angle~$\psi$ at which the slider starts to slip was measured.
From this the static friction coefficient can be determined via the relation~$\mu_\mrm{s}=\mrm{tan}(\psi)$.
The lower sliding part (and the additional weight) was removed for these tests in order to exclude the effect of wedging.
This analysis was repeated for ten equidistant starting positions along the beam's half in which the slider is supposed to move during the self-adaptive process.
The entire procedure was repeated three times (30 tests in total).
The resulting coefficient is in the range~$\mu_\mrm{s}\in[0.22, \ 0.31]$ with a mean value of 0.26.
Unavoidable imperfections of the beam surface stemming from the manufacturing process may be one reason for the variance.
However, variance is observed even for nominally identical test with the same starting position.
The coefficient of sliding friction is expected to be about one third lower than the static one for steel-steel combinations~(see \eg~\cite{Thomsen.2021}).
Coulomb's law assumes the same coefficient for sticking and sliding friction, which \MOD{is approximated by~$\mu=0.2\approx(\mu_\mrm{s}+2\mu_\mrm{s}/3)/2$ here}.
The effect of~$\mu$ on the results is discussed in~\sref{sensitivity}.
\\
\MOD{As in~\cite{Krack.2017,Muller.2020}, $\varepsilon_\mrm{n}=0.5$ is used as normal coefficient of restitution}. A parameter study showed that varying~$\varepsilon_\mrm{n}$ in the range $0.4 \leq \varepsilon_\mrm{n} \leq 0.6$ has no significant effect on the simulation results.

\subsection{Recap of time scale separation, super-slow manifold, system with pseudo-constrained slider, explanation of amplitude jumps}\label{sec:PCS_model}
Both experimental and numerical results show that the self-adaption process of the beam-slider system happens on \textit{three well-separated time scales}:
The beam's vibration takes place on the \textit{fast} time scale.
Modulations of the beam's vibration take place on an intermediate \textit{slow} time scale.
The adaption of the slider position as well as the change of the overall vibration amplitude take place on a \textit{super-slow} time scale.
By constraining the horizontal slider position and determining the steady-state vibration, \MOD{a subspace is obtained,} that \MOD{will be referred} to as \emph{super-slow manifold}.
The transient dynamics of the self-adaptive system with free slider closely follows the super-slow manifold.
The time scale separation and the notion of the super-slow manifold are crucial for the interpretation of the intricate nonlinear dynamics discussed in \sref{results}.
Before \MOD{proceeding} to the confrontation of experimental and simulation results, it is useful to describe how the super-slow manifold is computed and explain its characteristic shape.
\\
To directly compute the super-slow manifold, \MOD{the horizontal slider position needs to be treated} as a control parameter rather than as a degree of freedom.
In the past, \MOD{this was achieved} simply by prescribing the horizontal displacement of the slider's center of mass (\ie, imposing a vertical guidance).
As it turned out, this is too intrusive in the sense that the mean horizontal reaction force of the guidance was inconsistent with the actual movement of the slider in some parameter regimes.
Instead, \MOD{the slider is let} horizontally free, but \MOD{the prescribed (relative) slider position is always fed} to the algorithm that evaluates the contact kinematics.
This can be interpreted as moving the beam horizontally according to the slider's motion.
\MOD{This is referred to as \textit{model with pseudo-constrained slider} (PCS model) in the following}, which is to be distinguished from the model with free slider (FS model).
In \sref{results}, the super-slow manifold is computed using the PCS model and plotted to interpret the dynamics.
\begin{figure}
	\begin{center}
		\includegraphics[width=0.49\textwidth]{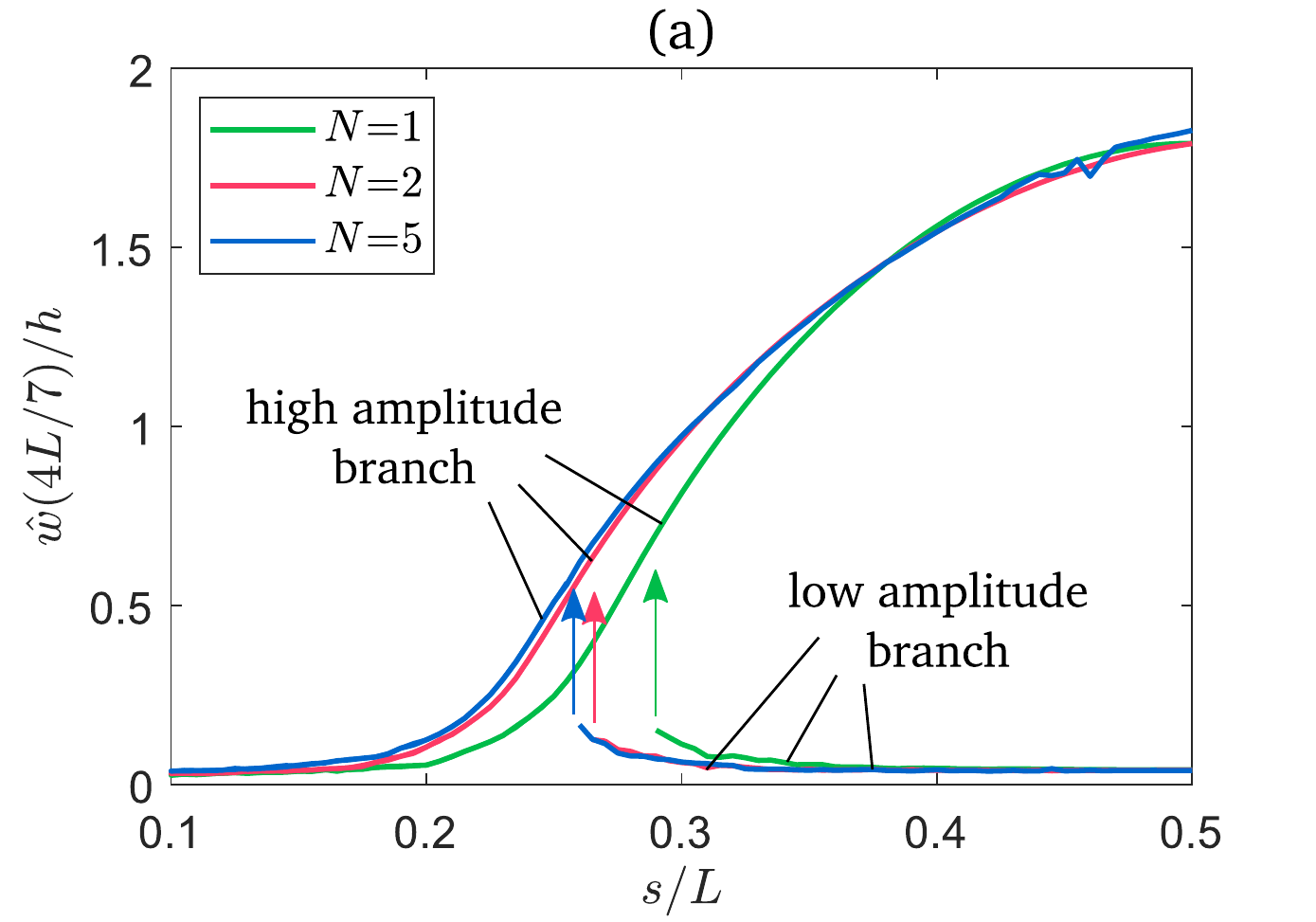}							\includegraphics[width=0.49\textwidth]{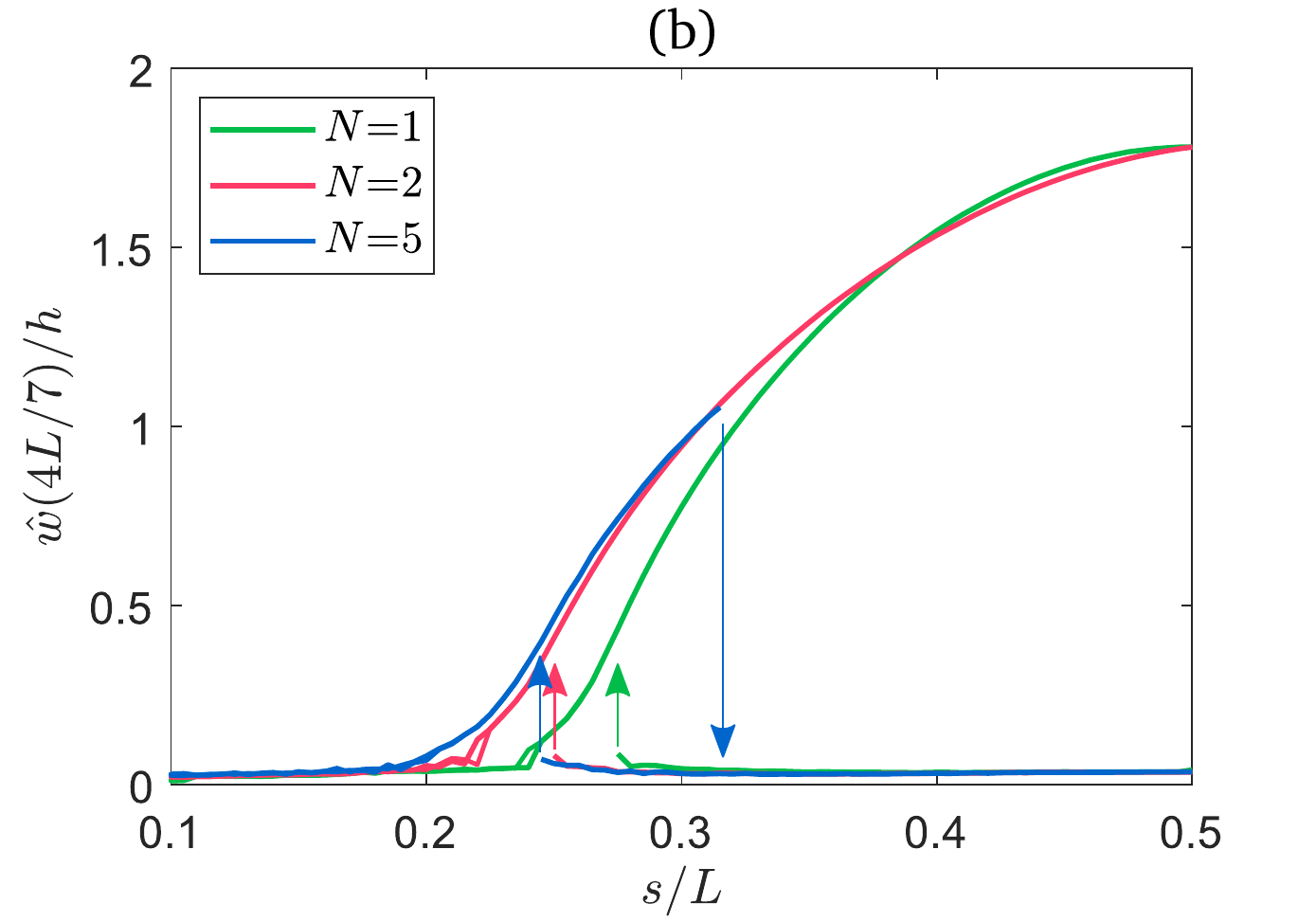}
		\caption{Bifurcation diagram illustrating the beam's response level under variation of slider position (super-slow manifold). PCS model evaluated with different modal truncation orders,~$f_\mathrm{ex}=124\mathrm{Hz}$ (a) excitation level~$\exa=14\mathrm{m/s}^2$ (b)~$\exa=\MOD{8}\mathrm{m/s}^2$}
		\label{fig:PCS}
	\end{center}
\end{figure}
\\
In \fref{PCS}, a typical super-slow manifold obtained using the PCS model is depicted in the plane spanned by the beam's vibration level and the slider position.
More specifically, the slider position, $s$, is normalized by the beam's length, $L$, and the beam's transversal displacement amplitude is determined at the location $4/7L$ and normalized by the beam's thickness.
Apparently, there are co-existing steady vibration states in a wide range of slider positions around the beam's center.
To resolve both stable branches of the super-slow manifold, \MOD{a sequential continuation procedure was used}:
The slider was initially placed near the beam's center (near the clamping), the simulation was run until steady state, and then the slider position is stepwise decreased (increased). 
Increments of $\Delta s/L={0.005}$ were found as reasonable compromise between resolution and effort, and a number of 300 excitation periods was found to ensure a reasonably steady state.
The upper (high-amplitude) branch extends from the clamping towards the beam's center.
The lower (low-amplitude) branch extends from the beam's center until it reaches a turning point, and thus a jump to the high-amplitude branch occurs during the sequential continuation towards the clamping.
\\
The super-slow manifold is depicted in \fref{PCS} with different modal truncation orders.
\fref{PCS}a suggests that $N=1$ is sufficient in the regions close to the clamping and the beam's center, whereas at least $N=2$ is needed to properly describe the intermediate range.
However, the results in \fref{PCS}b (lower excitation level) indicate that $N=2$ is not sufficient either:
The high amplitude branch corresponding to~$N=5$ terminates at~$s/L\approx0.32$ and the amplitude jumps down.
This behavior is crucial to understand one type of behavior of the system with free slider (relaxation oscillation) addressed in \sref{results}.
One possible explanation for the jump down may be a modal interaction which is not captured in models with~$N\leq 2$.
A more thorough analysis of this phenomenon is considered as beyond the scope of the present work. Further increasing the number of retained modes ($N>5$) does not significantly affect the results, as found in a preliminary study.
\section{Confronting experimental and simulation results}\label{sec:results}
In this section, \MOD{a validation of the \textit{model with horizontally free slider} (FS model) by means of the experiment is presented}.
Of course, the initial slider position is an important parameter, \MOD{which was varied} in a wide range.
In the present work, however, \MOD{results are shown for initial slider position $s/L=0.3$ (unless otherwise stated)}.
As will be shown, all types of behavior can then still be reached by varying the excitation frequency and excitation level (\emph{operating point}).
\MOD{Starting at static equilibrium, each} operating point is driven for 120 seconds which is sufficient to reach a reasonably steady state. \MOD{Representative results measured from the experiments are available in a repository \cite{Muller.2022c}.}
Four qualitatively different types of behavior are distinguished, as mentioned in the introduction:
A successful adaption to high vibrations in the form of the signature move,
the trivial adaption to high vibrations,
a relaxation oscillation and
self-damping behavior.
\MOD{Representative results for each type of behavior are shown} in the following Subsections \srefo{sign_m}-\srefo{Self-damping}.
A very good agreement of experiment and simulation is observed for all cases.
Minor deviations are explained by high sensitivity with respect to uncertain system parameters in \ssref{sensitivity}.
\MOD{A more global picture is presented in \ssref{op_map},
showing which type of behavior is reached for what operating conditions.}

\subsection{Successful adaption (signature move)}\label{sec:sign_m}
Results for a representative signature move are depicted in~\fref{sign_m}.
The process can be subdivided in three phases.
The transitions between these phases are indicated by vertical dashed lines in \fref{sign_m}a,b,e.
In phase~1, the vibration level is small and the slider moves away from the beam's center.
A closer look to the beam's transversal displacement during phase~1 shows a strongly modulated response (\fref{sign_m}c,d).
At the transition from phase~1 to phase~2, the amplitude jumps to a higher level, after which the slider moves back towards the beam's center and the amplitude increases further.
In phase~3, the horizontal slider position and large vibration level are maintained (steady state).
\\
The three separate time scales described in \ssref{PCS_model} can be well recognized in \fref{sign_m}.
The time history within phase~1 depicted in \fref{sign_m}c,d clearly shows the intermediate time scale, on which the amplitude modulation takes place, and the fast time scale of the oscillation.
The time variation of the horizontal slider location $s$, depicted in \fref{sign_m}e, happens on a much slower (super-slow) time scale.
Consequently, the transient dynamics of the model with free slider (FS model) closely follows the super-slow manifold corresponding to the steady-state dynamics of the model with pseudo-constrained slider (PCS model), see \fref{sign_m}f.
More specifically, the FS model initially follows the low-amplitude branch of the PCS model (phase~1).
The amplitude jumps precisely at the position predicted by the PCS model, and then the FS model follows the high-level branch (phase~2) until steady state is reached (phase~3).
To obtain the time-dependent super-slow amplitude of the beam's vibration (both from the experimental data and the simulation data of the FS model), \MOD{the Hilbert transform was used and the outcome was subjected to} a moving average with a window time of 1 second.
\\
\MOD{Model} and experiment are in excellent qualitative agreement with regard to all features of the signature move, including the strongly modulated character of the response in phase~1.
Quantitatively, the amplitude level and the speed of adaption (\ie the speed of the slider's horizontal movement) are in excellent agreement for phase~1.
Additionally, the speed of adaption is very similar in phase~2.
\MOD{The slider's turning point that separates phase~1 and 2 as well as the final slider position are} slightly closer to the beam's center in the experiment.
A possible explanation for \MOD{the deviation of the final slider position} is the uncertainty of the friction coefficient (systematic deviation between static vs. dynamic friction coefficient, deterministic variation along the beam's length, plus aleatoric uncertainty).
This inevitably leads to some deviation, for instance in case one of the states (sticking or sliding) dominates.
The high sensitivity of the final slider position with respect to the friction coefficient is analyzed in~\sref{sensitivity}.
The (averaged) steady-state amplitude is almost identical for experiment and simulation, which one can see in  \fref{sign_m}f.
In that sub-figure, it can also be seen that the measured super-slow trajectory is very close to the simulated one, merely shifted slightly towards the right. \MOD{A second example is depicted in~\fref{var_start_pos}b, illustrating that the signature move is also possible in case the slider is placed closer to center initially.}
\begin{figure}
	\begin{center}
		\includegraphics[width=0.49\textwidth]{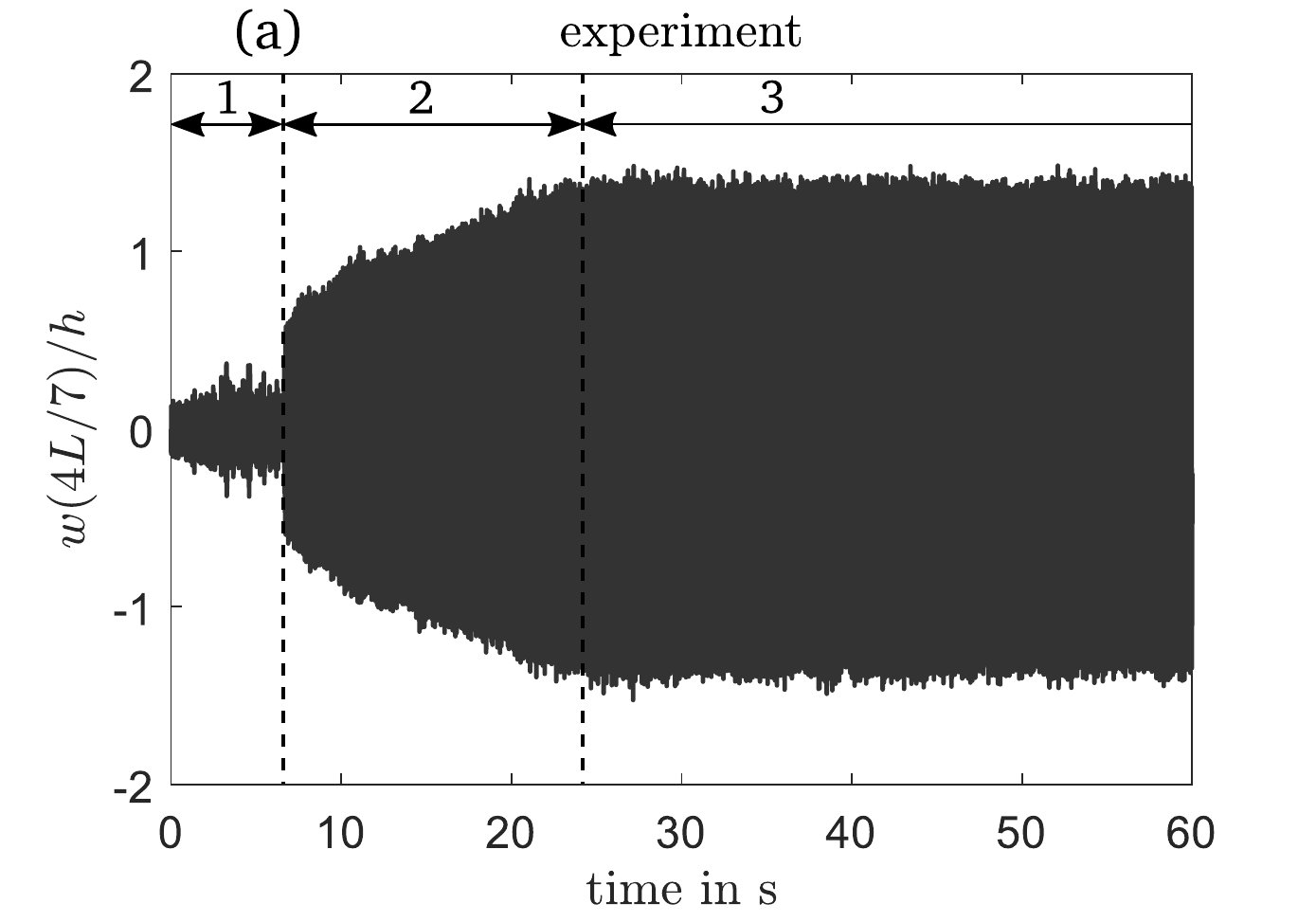}
		\includegraphics[width=0.49\textwidth]{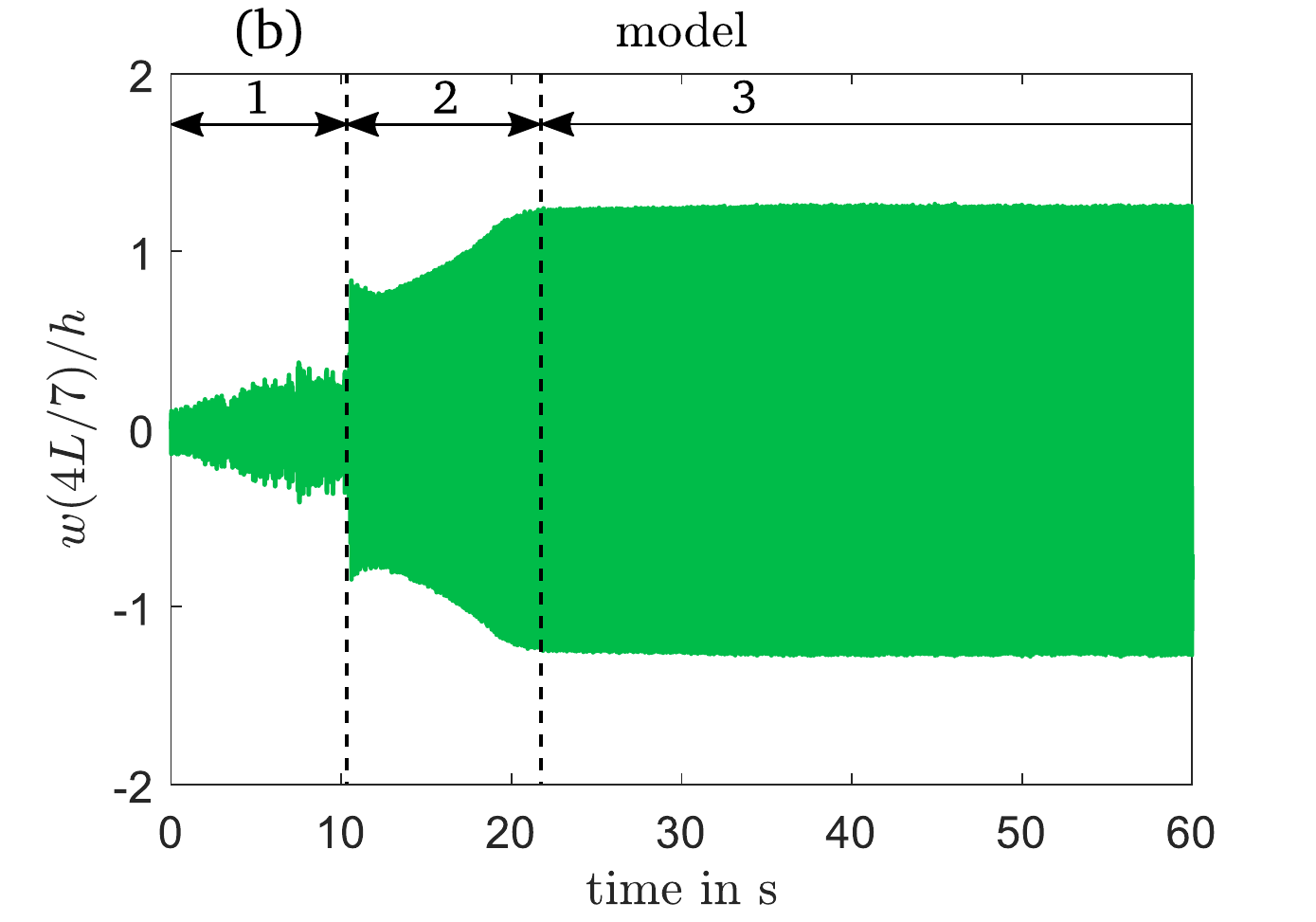}
		\includegraphics[width=0.49\textwidth]{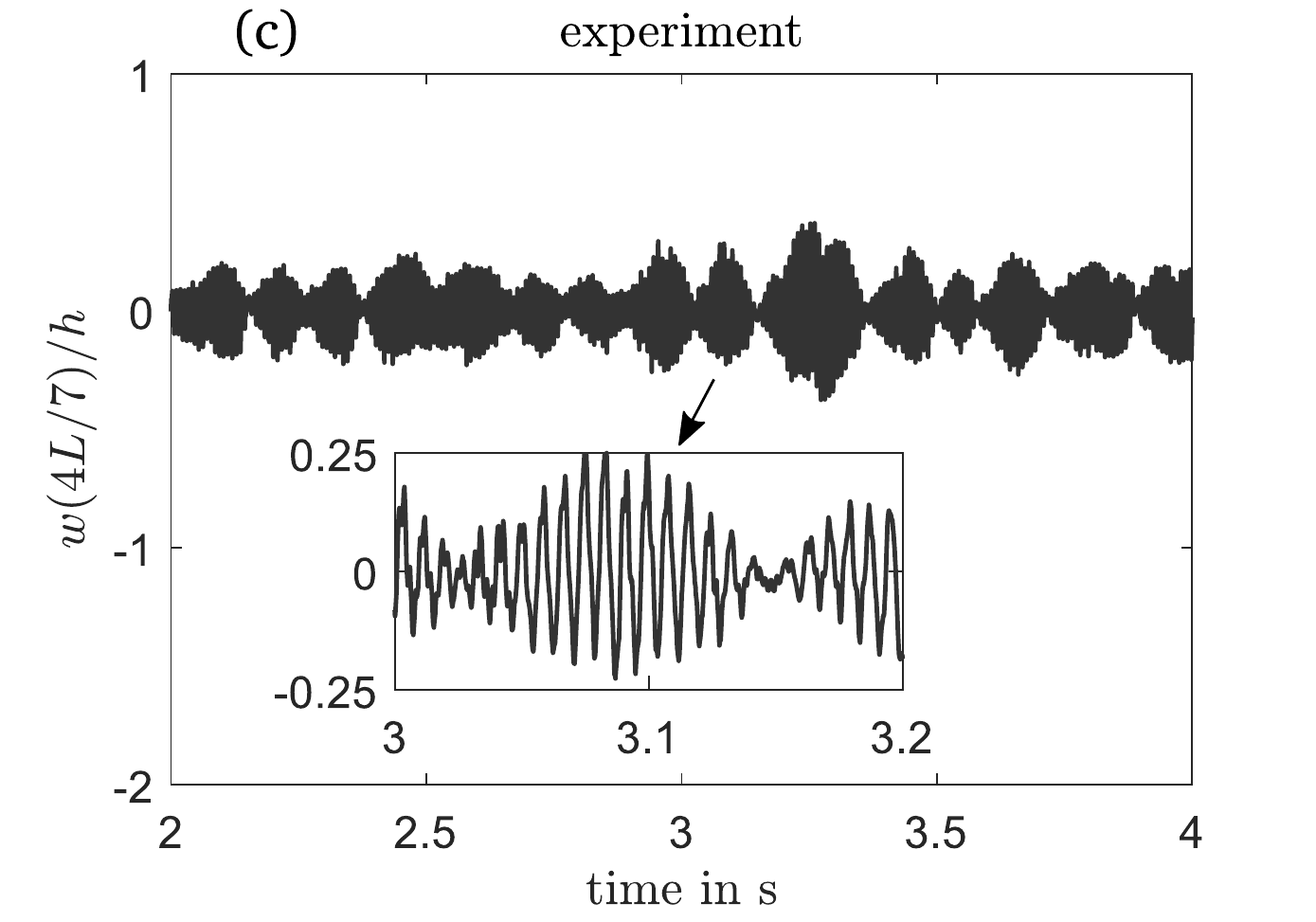}
		\includegraphics[width=0.49\textwidth]{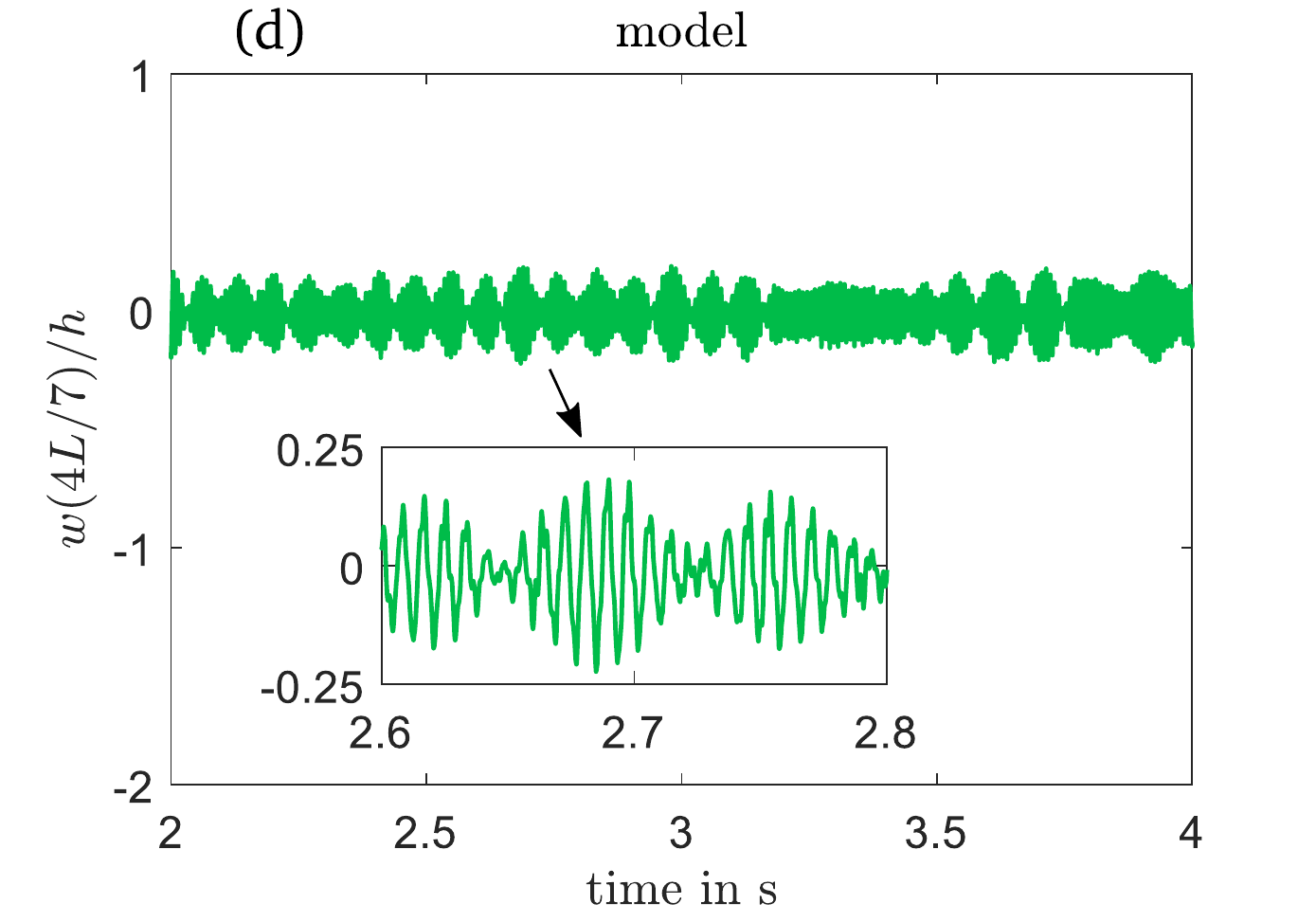}		
		\includegraphics[width=0.49\textwidth]{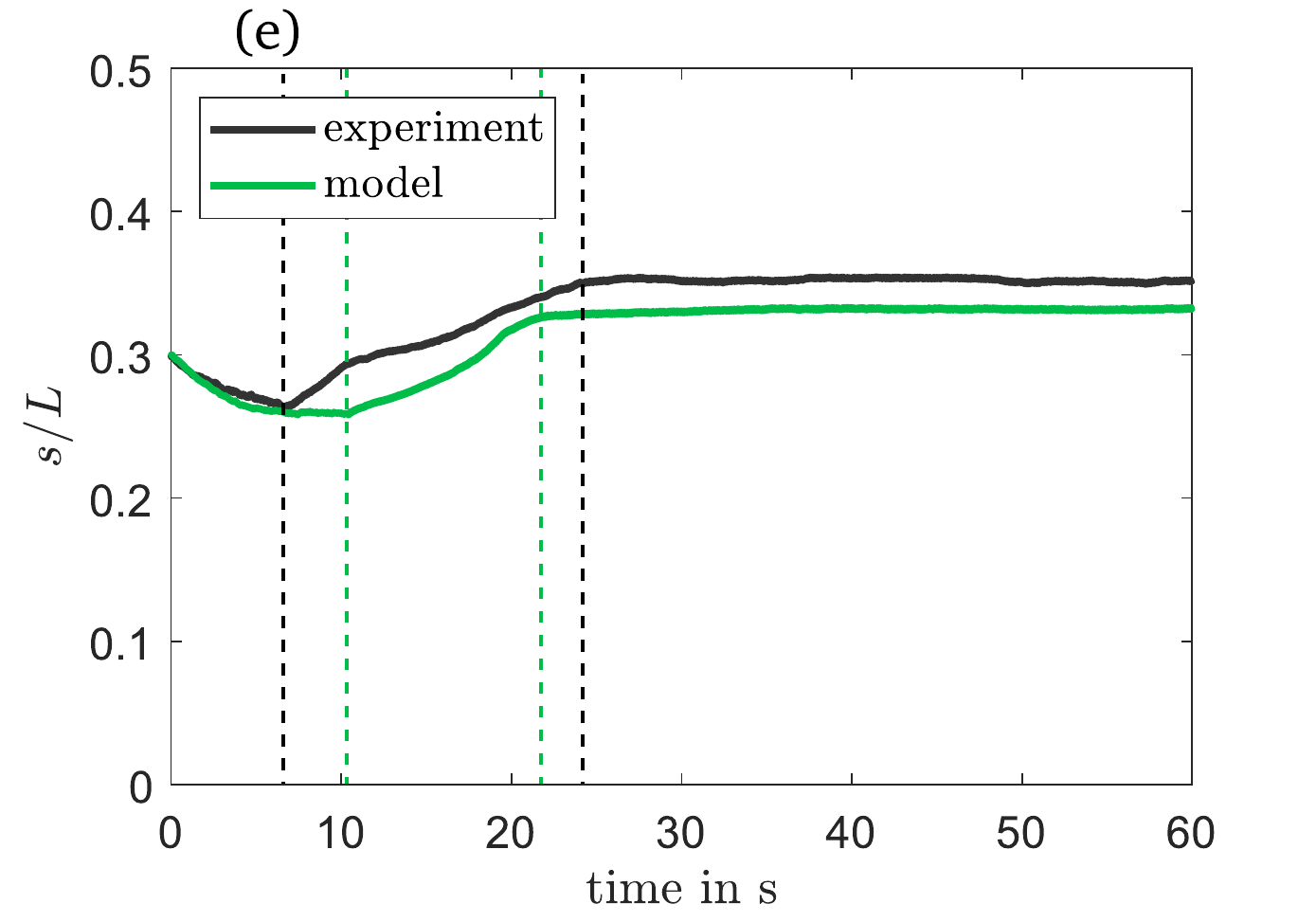}
		\includegraphics[width=0.49\textwidth]{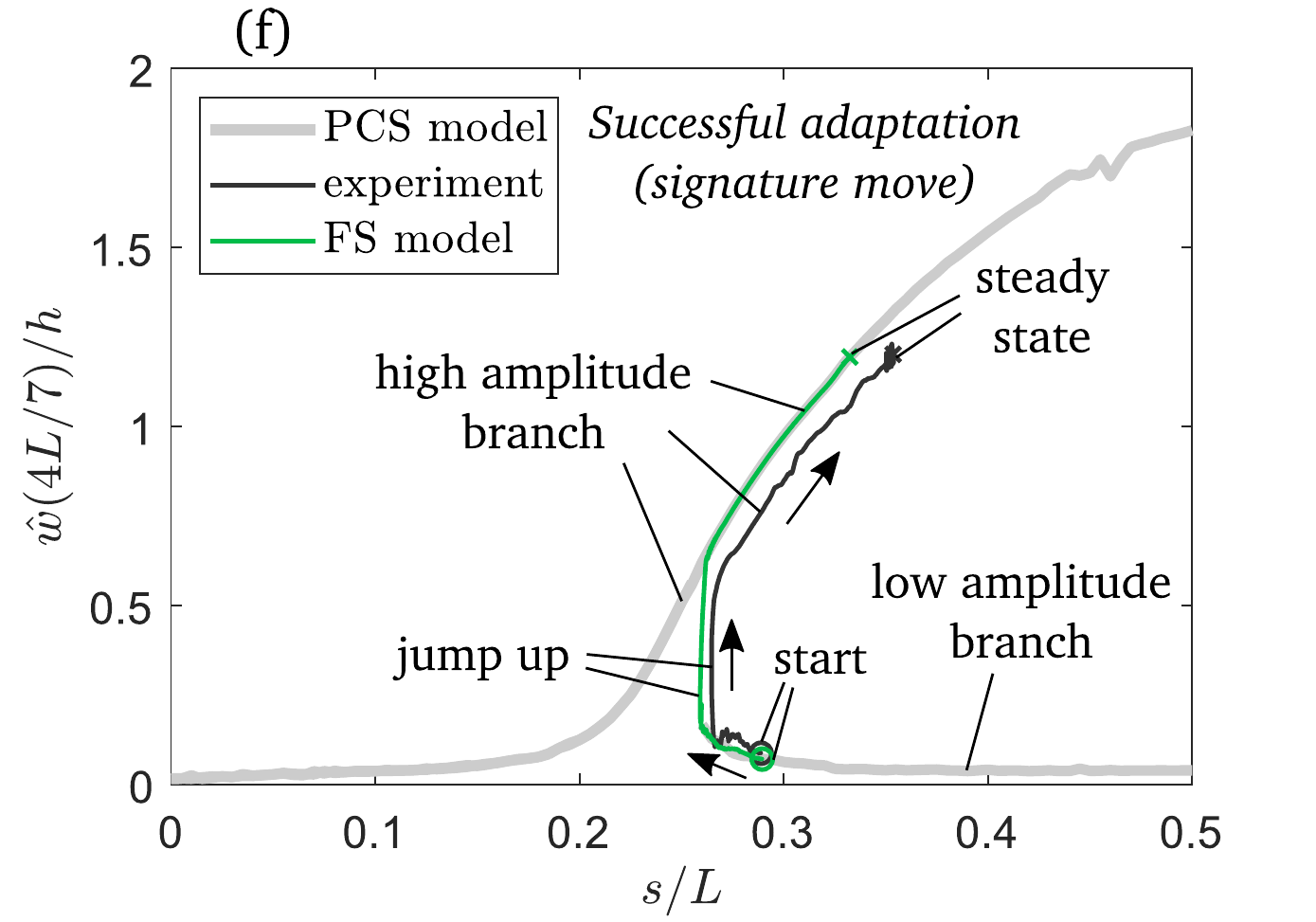}
		\caption{Successful adaption (signature move): Results of experiment and simulation 				model at~$f_\mathrm{ex}=124\mathrm{Hz}$,~$\exa=14\mathrm{m/s}^2$ (a)-(d) beam's deformation vs. time (e) slider position vs. time (f) super slow dynamics}
		\label{fig:sign_m}
	\end{center}
\end{figure}
\subsection{Successful adaption (trivial)}\label{sec:trivial}
A representative example of trivial successful adaption is depicted in~\fref{trivial}.
The excitation frequency is significantly lower compared to the signature move example.
In the experiment, the slider moves monotonously towards the beam's center first (phase~1) and then assumes a certain steady-state position (phase~2), see \fref{trivial}c,d.
The beam's vibration level increases during phase~1 and is maintained at high level in phase~2~(\fref{trivial}a,b).
This type of behavior is also captured well by the simulation.
Minor deviations exist in the the final vibration level and the time history of the slider position.
As for the signature move, there is only a minor offset between the super-slow trajectories of experiment and FS model.
\begin{figure}
	\begin{center}
		\includegraphics[width=0.49\textwidth]{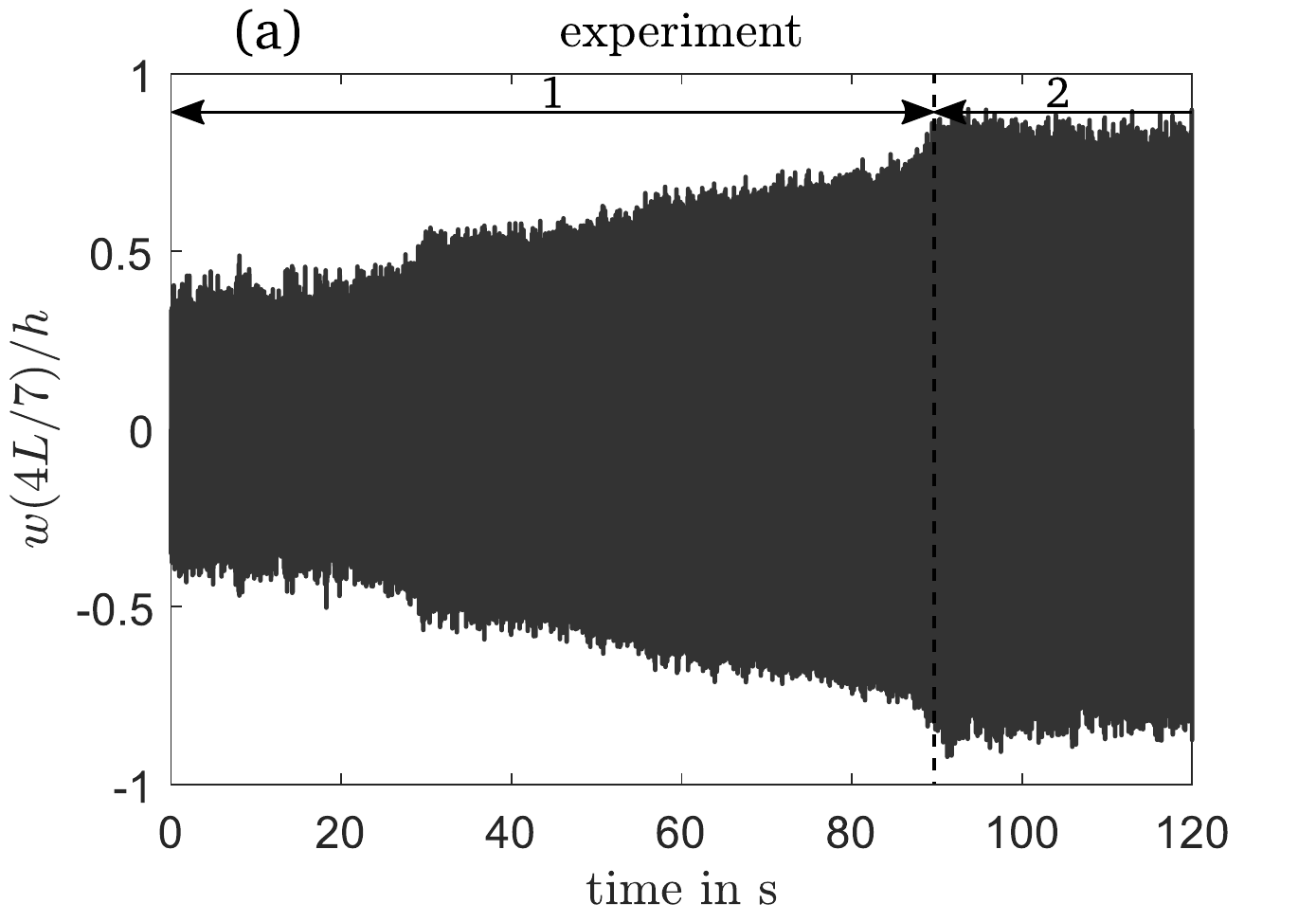}
		\includegraphics[width=0.49\textwidth]{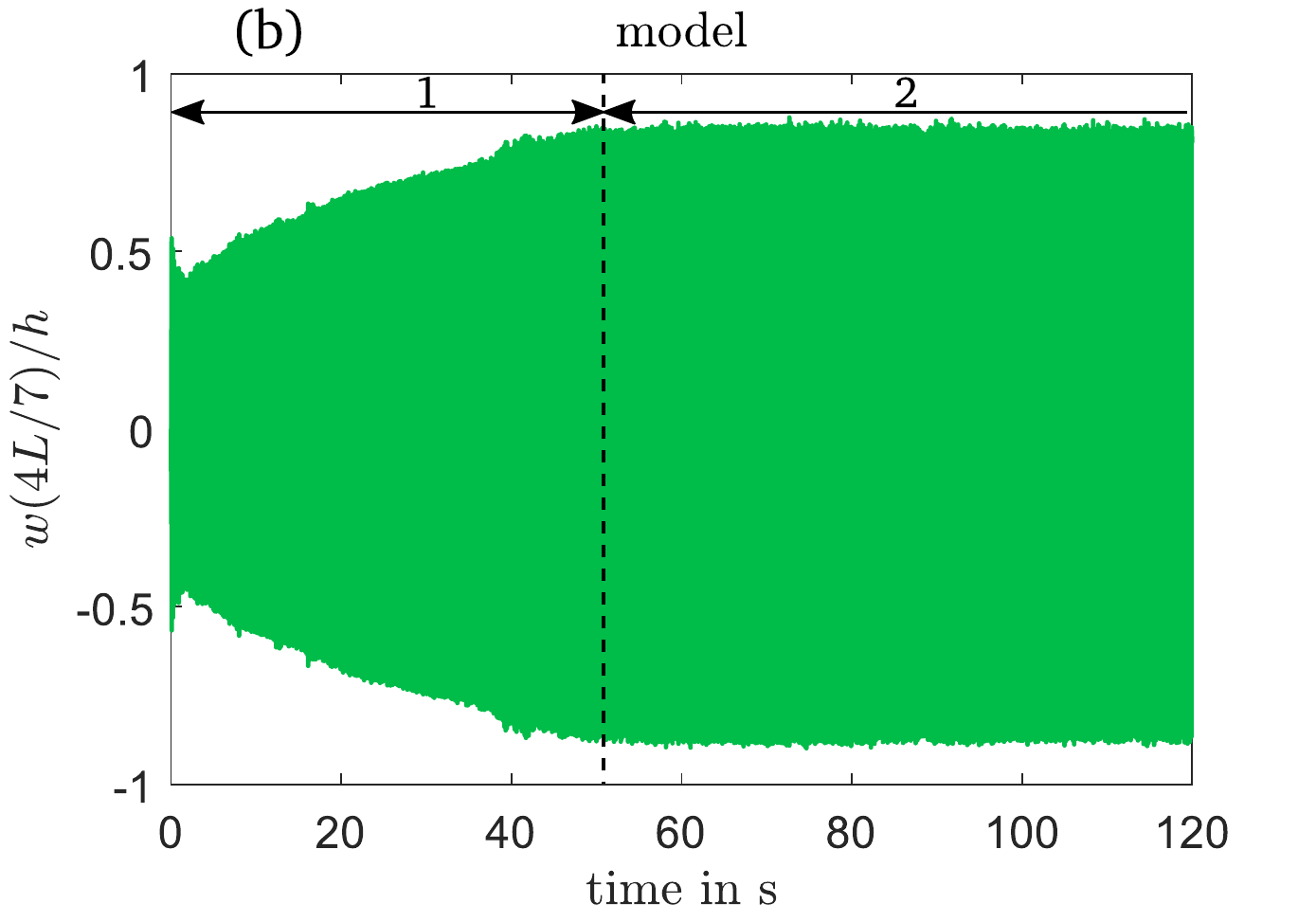}
		\includegraphics[width=0.49\textwidth]{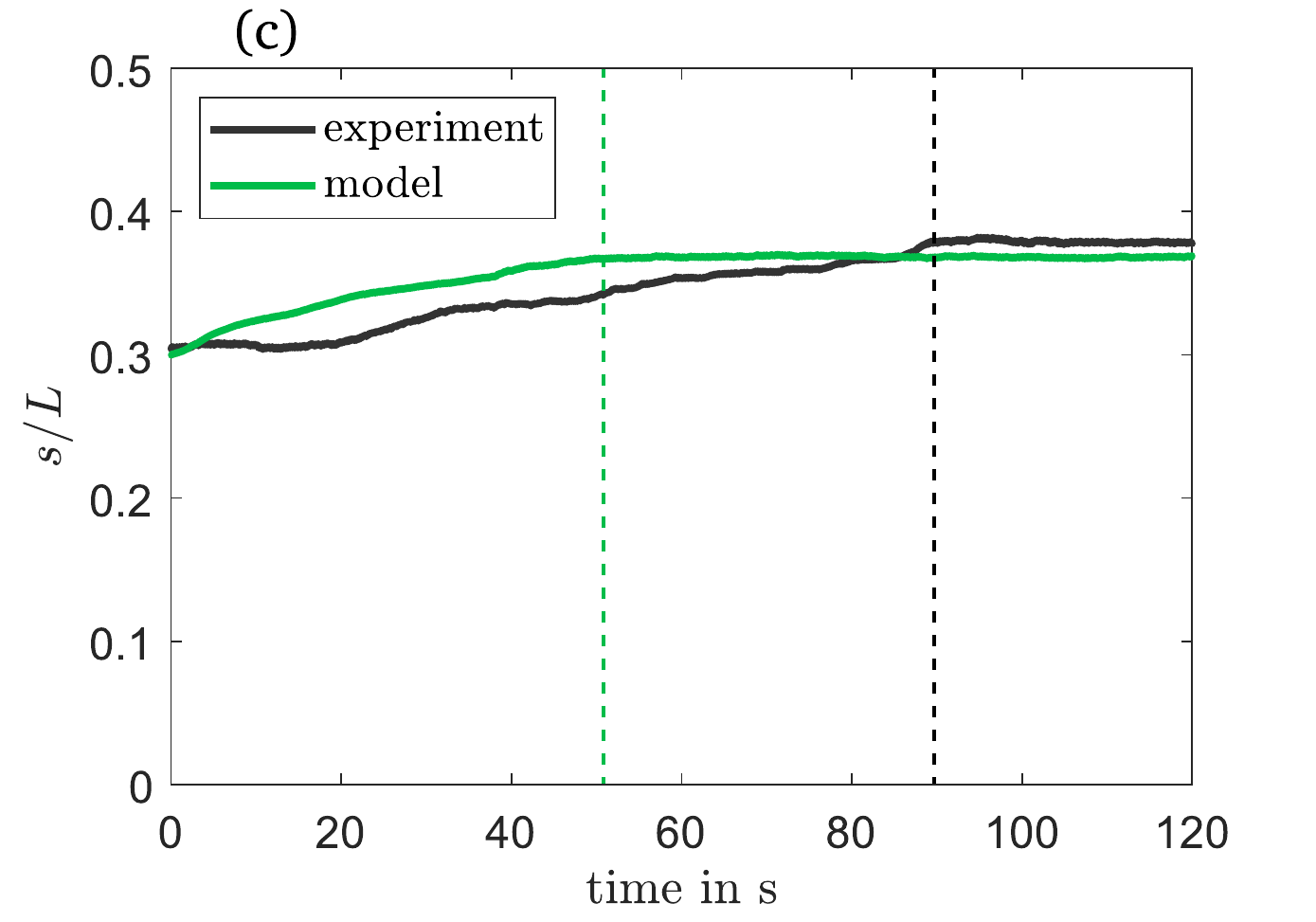}
		\includegraphics[width=0.49\textwidth]{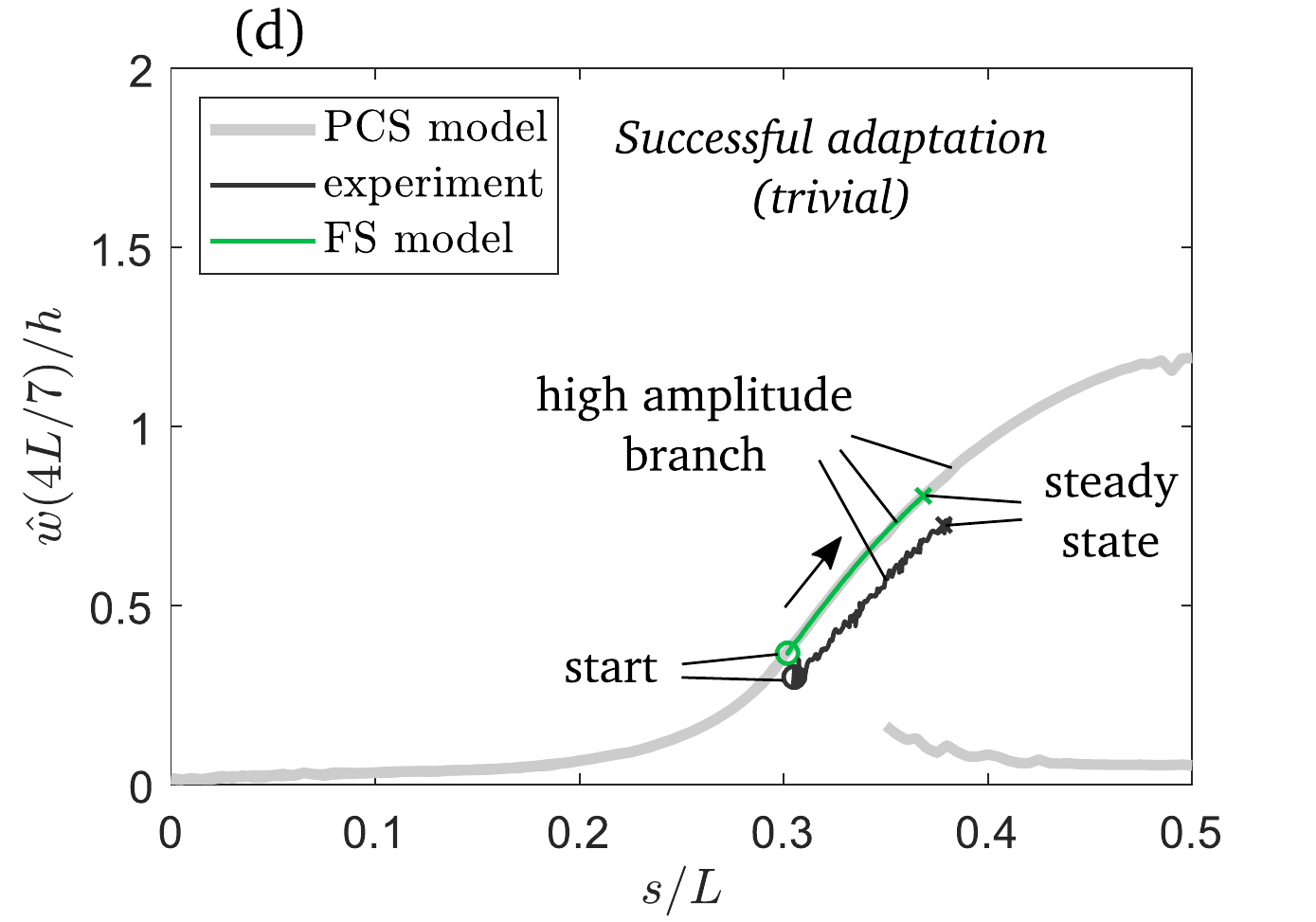}
		\caption{Successful adaption (trivial): Results of experiment and simulation model at~$f_\mathrm{ex}=104\mathrm{Hz}$,~$\exa=14\mathrm{m/s}^2$ (a)-(b) beam's deformation vs. time (c) slider position vs. time (d) super slow dynamics}
		\label{fig:trivial}
	\end{center}
\end{figure}
\subsection{Relaxation oscillation}\label{sec:relax}
A representative case of a relaxation oscillation is shown in~\fref{relax}.
Phase~1 and~2 are similar to the first two phases of the signature move, however, phase~2 ends with a sudden jump down to the low-amplitude branch.
After this, the process repeats approximately periodically, yielding a strongly modulated response in terms of the beam's vibration on the super-slow time scale.
Also this highly complex type of behavior is well-reproduced by the model.
Quantitative features, such as the amplitude and the duration of each super-slow oscillation cycle are in very good agreement.
As already discussed in~\sref{PCS_model}, a sufficient number of modes ($N\geq5$) needs to be retained in the modal beam model to predict the amplitude jump down.
\begin{figure}
	\begin{center}
		\includegraphics[width=0.49\textwidth]{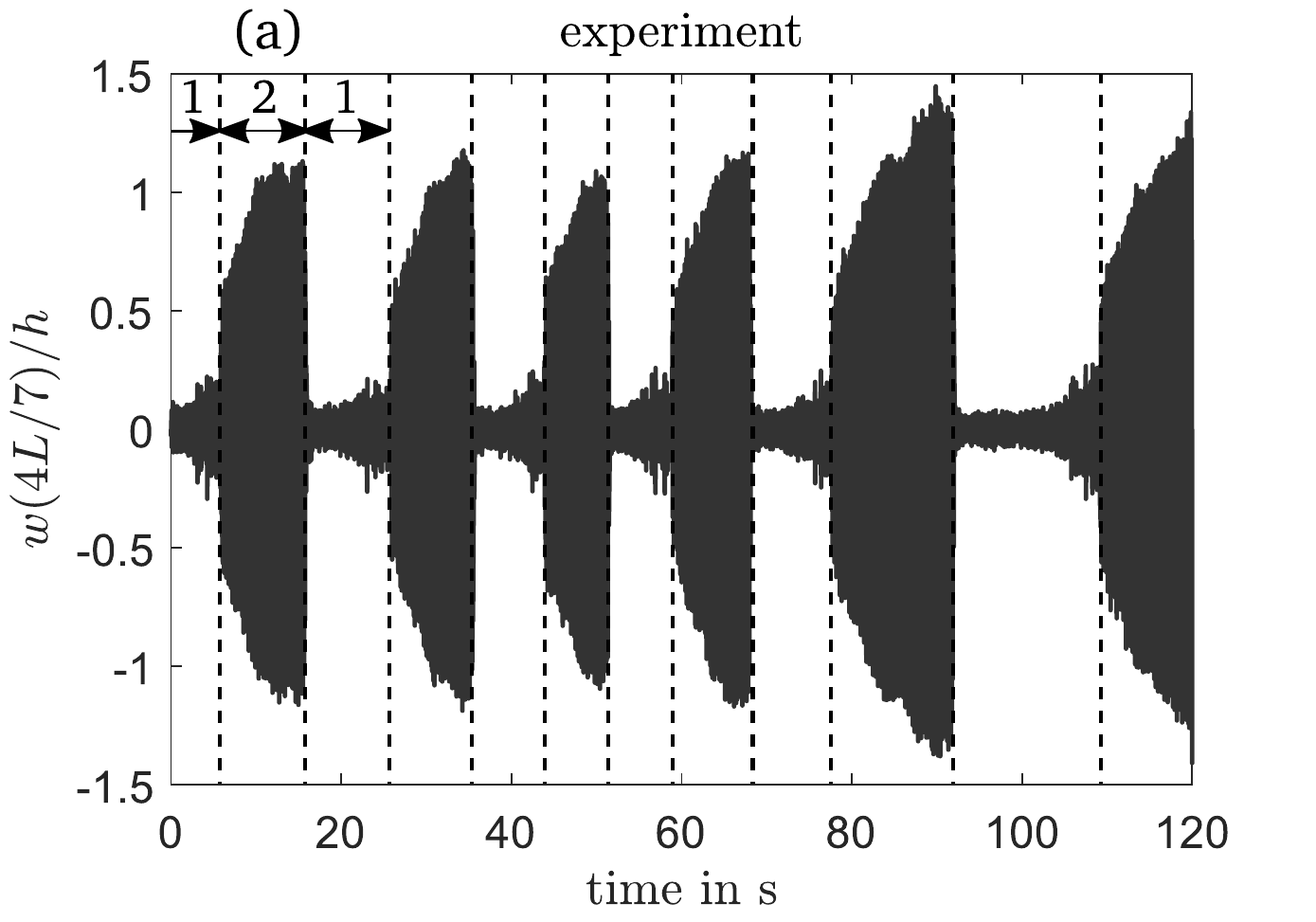}
		\includegraphics[width=0.49\textwidth]{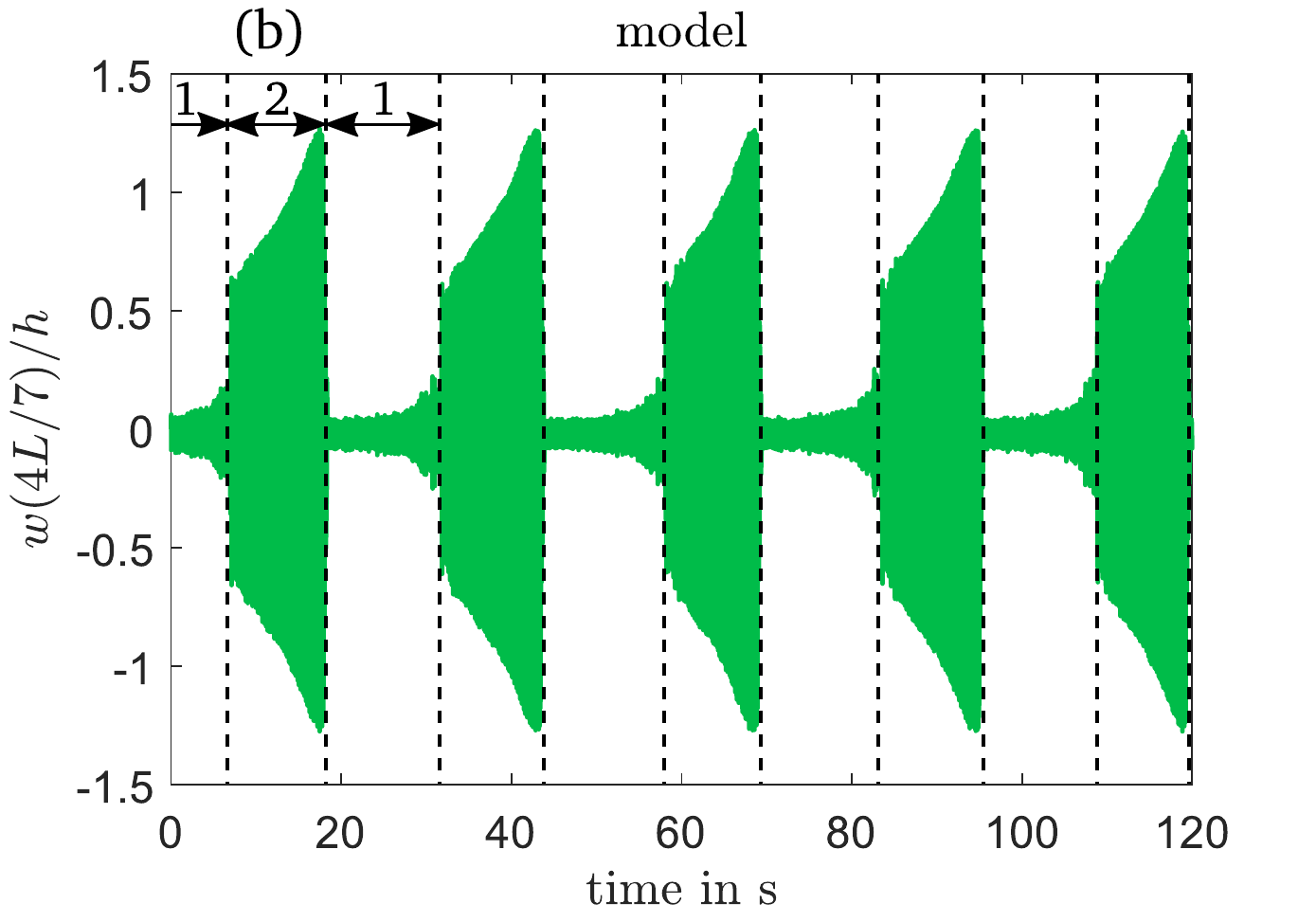}
		\includegraphics[width=0.49\textwidth]{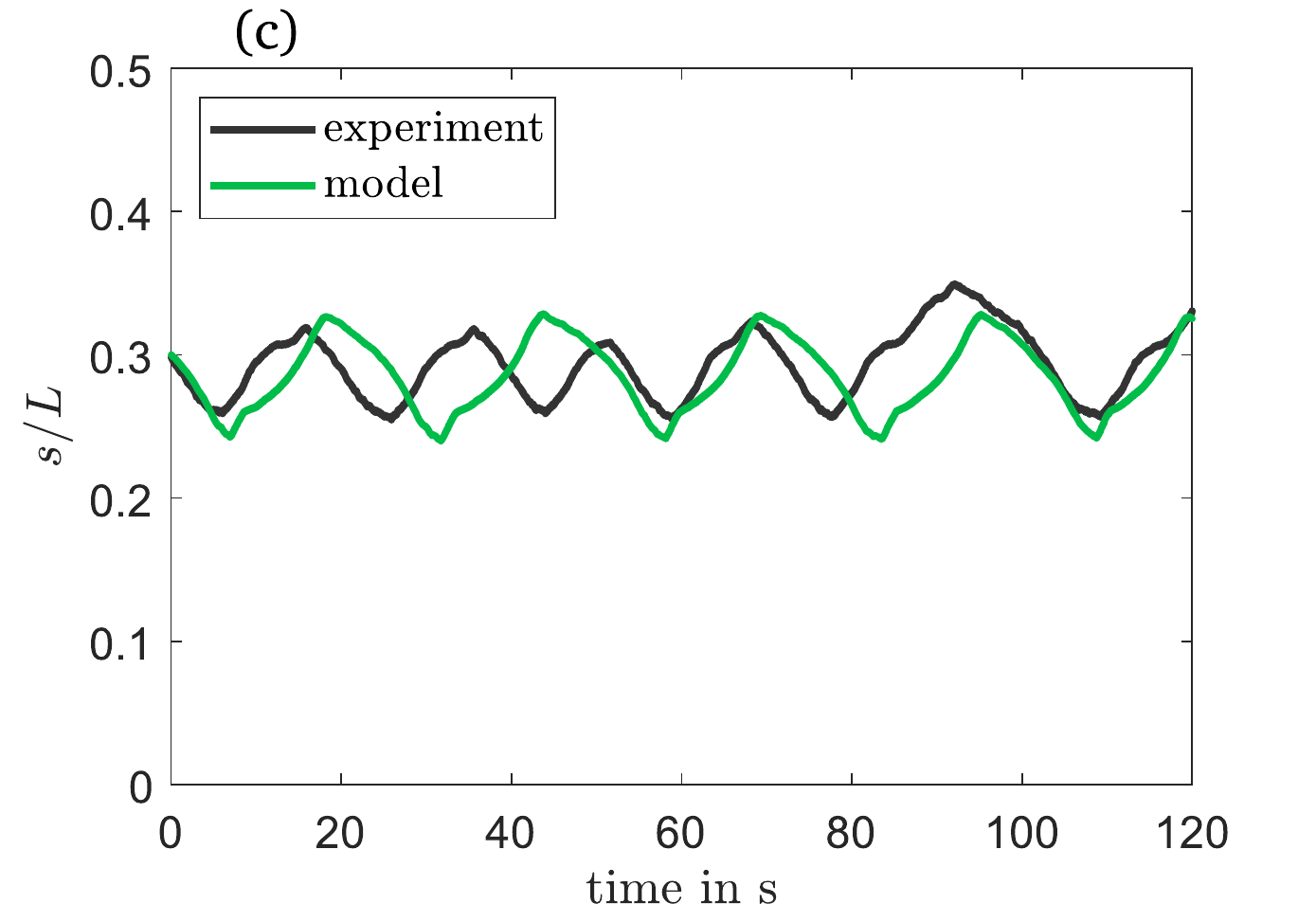}
		\includegraphics[width=0.49\textwidth]{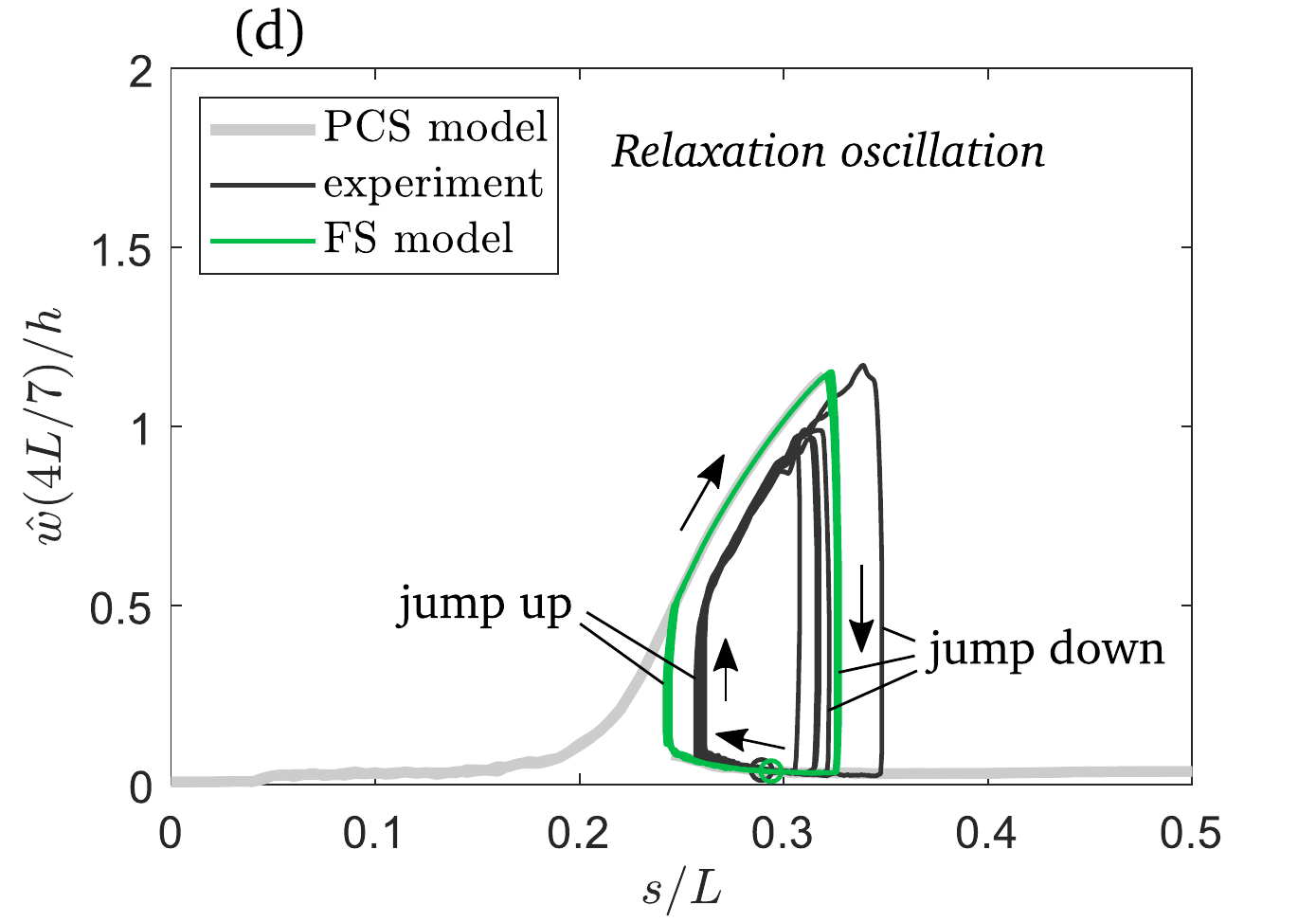}
		\caption{Relaxation oscillation: Results of experiment and simulation model at~$f_\mathrm{ex}=\MOD{126}\mathrm{Hz}$,~$\exa=10\mathrm{m/s}^2$ (a)-(b) beam's deformation vs. time (c) slider position vs. time (d) super slow dynamics}
		\label{fig:relax}
	\end{center}
\end{figure}
\subsection{Self-damping}\label{sec:Self-damping}
The last \MOD{presented type of behavior} is self-damping, see~\fref{self-damping}.
Here, the excitation frequency is the same as in the case of the signature move, but the excitation level is significantly decreased.
Phase~1 again is similar to the signature move; \ie, the slider moves away from the beam's center and the vibration amplitude increases slightly.
After the jump to the high-amplitude branch, however, the slider does not turn.
Instead, it keeps moving towards the clamping and, in accordance with the PCS model, the amplitude decreases again (phase~2).
At the end of phase~2 and during phase~3, the amplitude is so small that the upper contact points do not lift off anymore.
This can be easily verified in the simulation.
Consistently, in the experiment, the rattling noise abruptly stops at the beginning of phase~3.
Subsequently, the beam vibrates harmonically at very low level and the slider's position is maintained at~$s/L\approx0.1$.
Again, the simulation results agree excellently with the measurements.
The amplitude jumping is less clear in the experimental results than in the simulated ones.
In fact, several jumps up and down are observed in the experiment.
This is attributed to the poor performance of the excitation level control in that time span.
The controller is relatively slow so that it was unable to maintain a constant excitation amplitude.
More specifically, the excitation amplitude varied by $\pm1.6\mathrm{m/s}^2$ (nominal value $6\mathrm{m/s}^2$) in that time span. \MOD{In the simulation, the amplitude jump is smaller compared to the experiment. In the following subsection it is shown that this deviation can be explained by the high sensitivity of the process with respect to uncertain model parameters.}
\begin{figure}
	\begin{center}
		\includegraphics[width=0.49\textwidth]{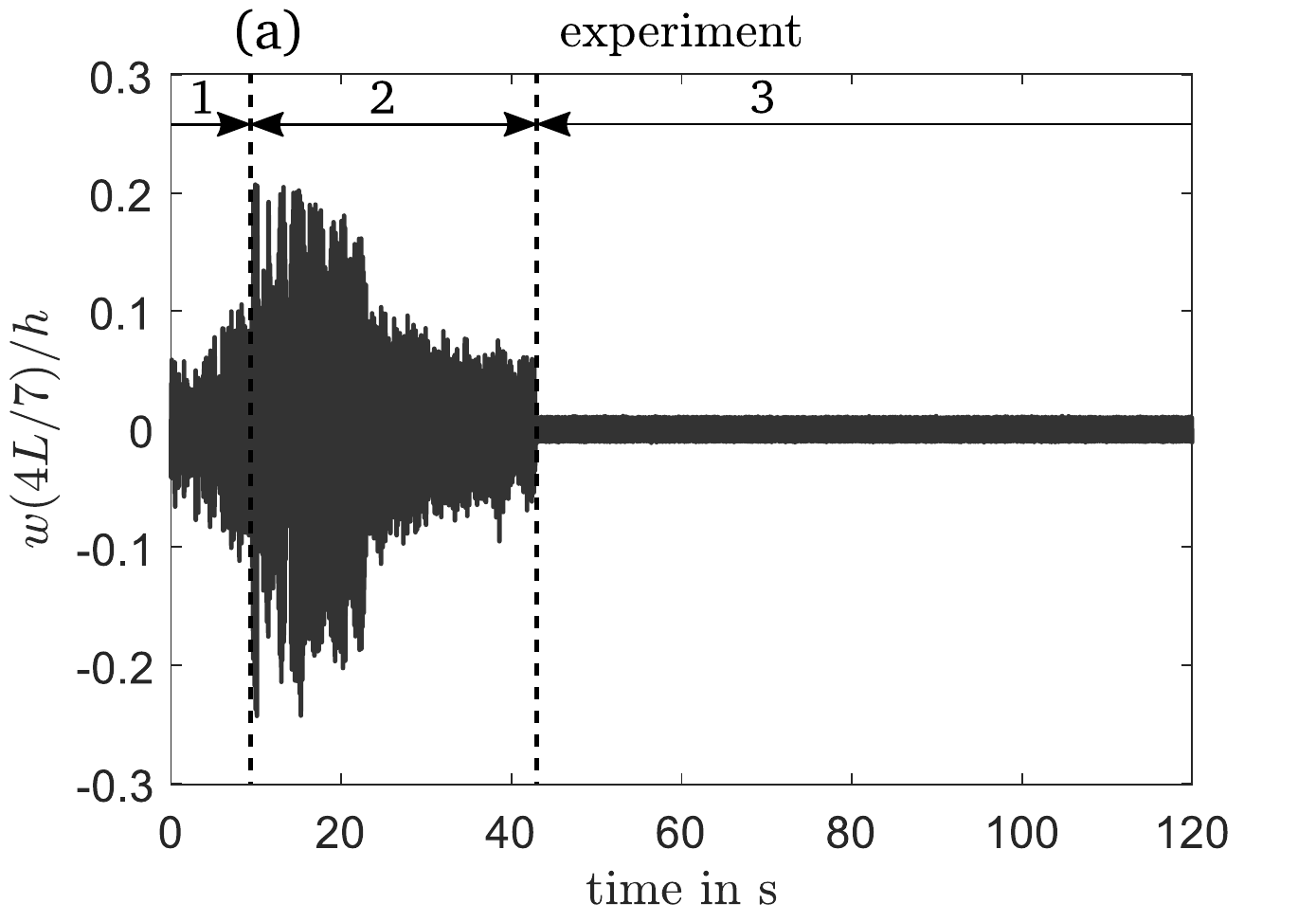}
		\includegraphics[width=0.49\textwidth]{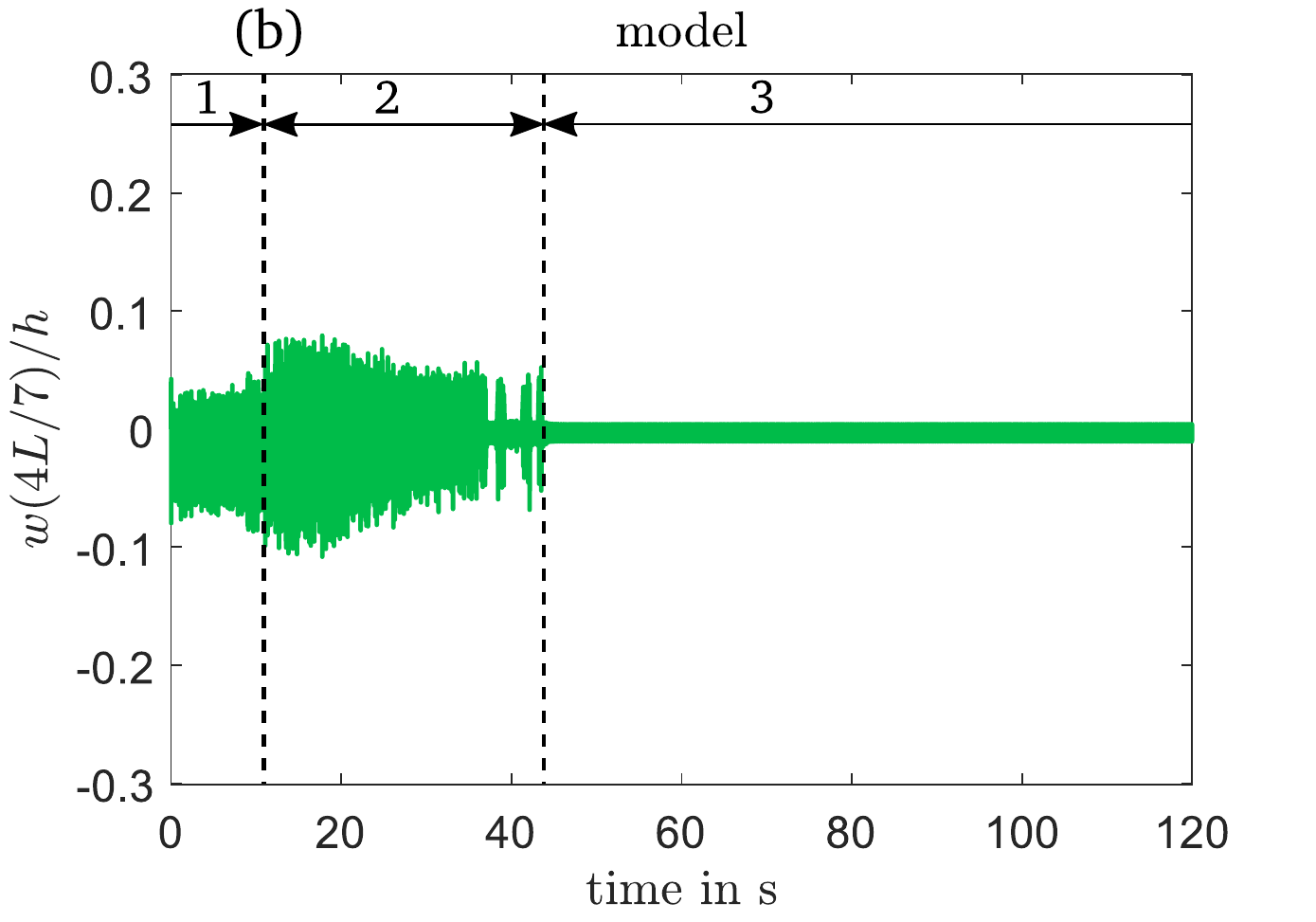}
		\includegraphics[width=0.49\textwidth]{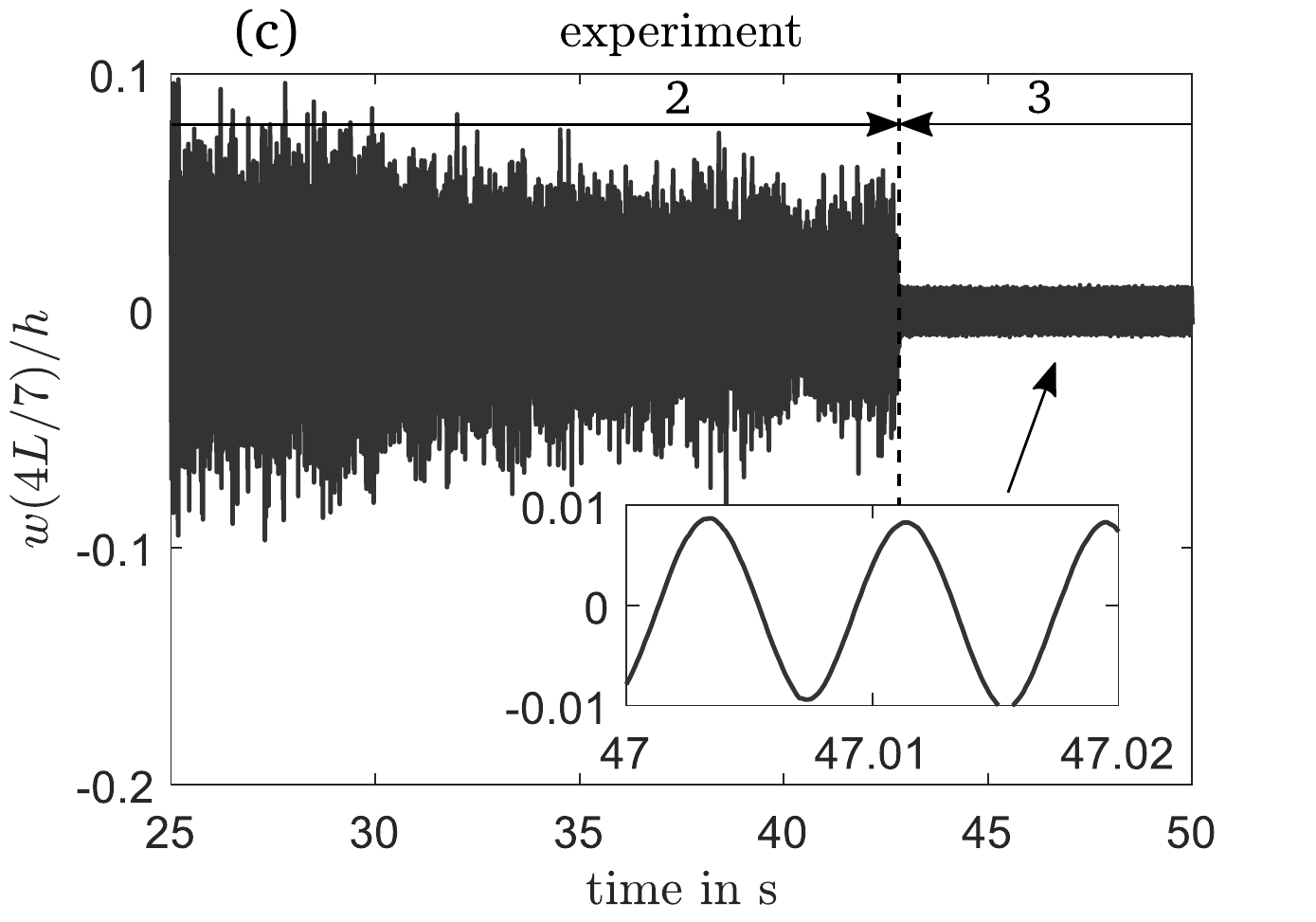}
		\includegraphics[width=0.49\textwidth]{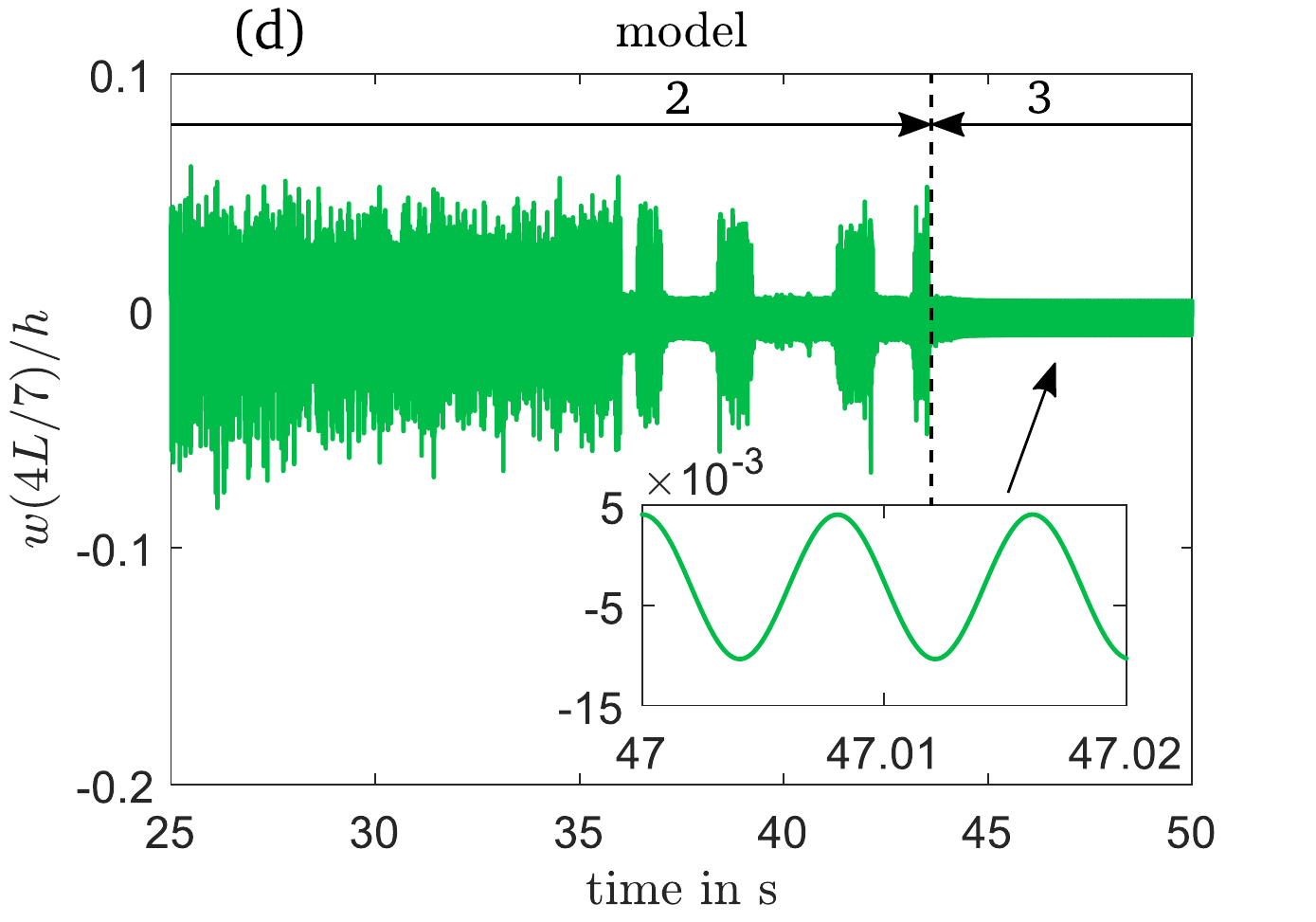}
		\includegraphics[width=0.49\textwidth]{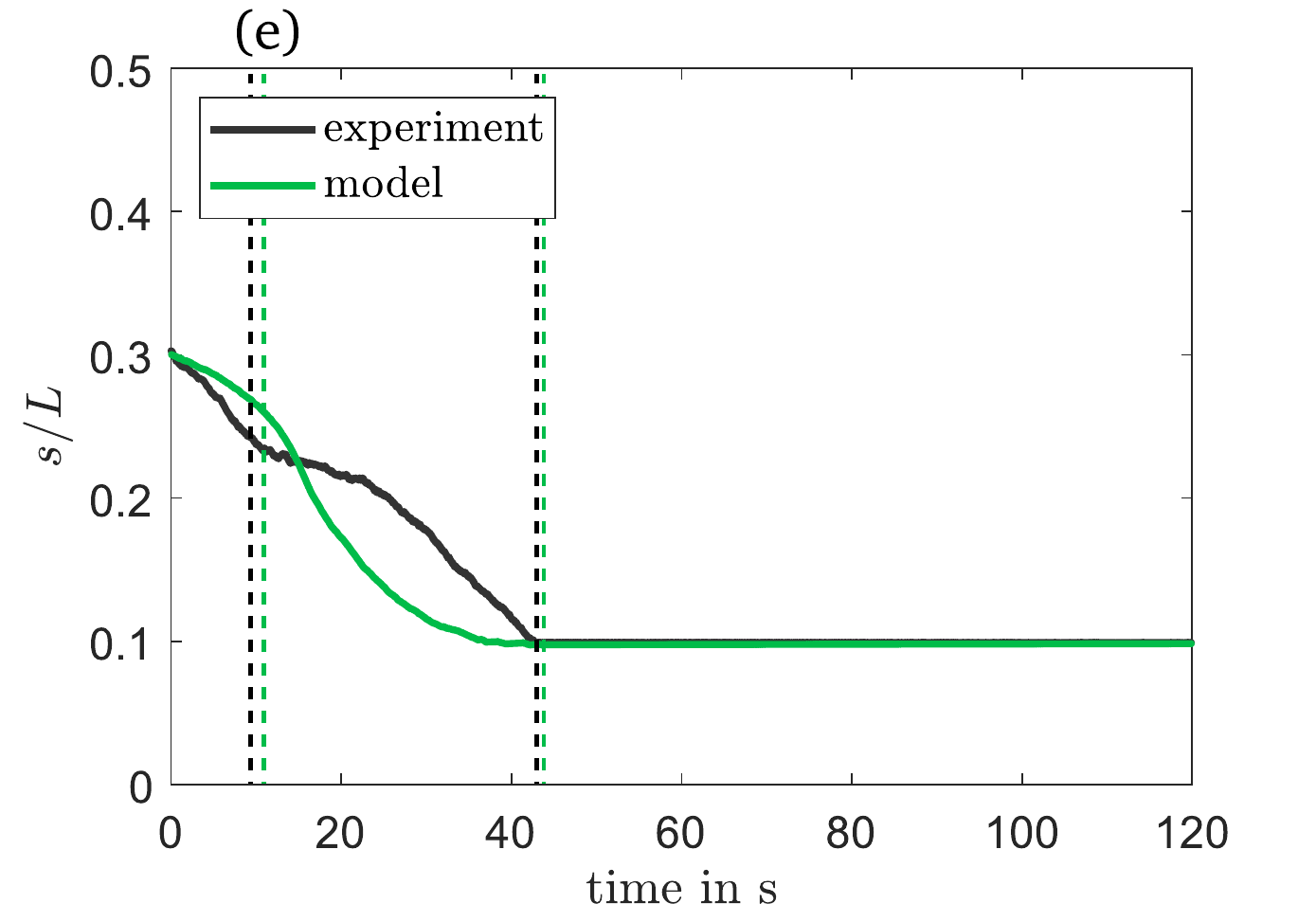}
		\includegraphics[width=0.49\textwidth]{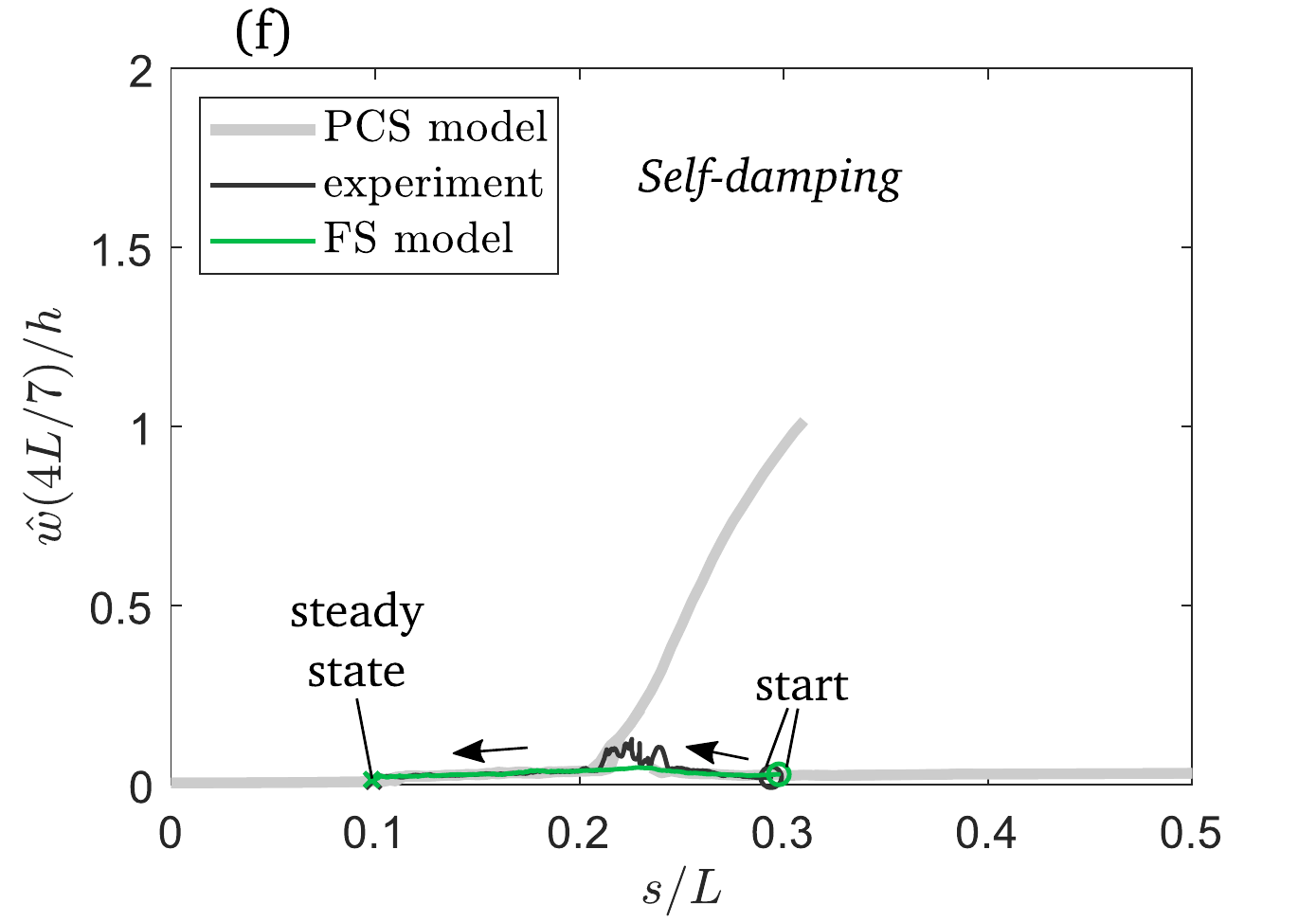}
		\caption{Self-damping: Results of experiment and simulation model at~$f_\mathrm{ex}=124\mathrm{Hz}$,~$\exa=6\mathrm{m/s}^2$ (a)-(d) beam's deformation vs. time (e) slider position vs. time (f) super slow dynamics
}
		\label{fig:self-damping}
	\end{center}
\end{figure}
\subsection{Sensitivity with respect to experimental imperfections}\label{sec:sensitivity}
In conclusion, all different types of behavior that are observed in the experiments can be reproduced by the model in an excellent manner.
Only minor quantitative deviations of the super-slow dynamics, more particular the steady slider position and the amplitude--slider position trajectory are encountered.
The small shift of this trajectory may be explained by the fact that the experiments with attached slider and the ENMA defining the beam model could not be done on the same day due to the large time required for setting up the instrumentation for the ENMA and varying the excitation parameters in wide ranges.
Moreover, beam and frame needed to be disassembled and reassembled in between the ENMA of the beam and the study of the beam-slider system.
Thus, a certain deviation of the linear natural frequency is to be expected.
To show how sensitive the super-slow manifold is with respect to small variations of the updated beam model, \MOD{the lowest linear natural frequency is deliberately de-tuned} by~$\omega^*=\omega\pm 5\%$.
This results in a shift of the turning point by~$\Delta s/L\approx{\pm} 0.013$ and maximum amplitude~$\Delta\hat{w}(4L/7)/h\approx{\mp}\MOD{0.088}$, as illustrated in \fref{sensitivity}a.
\begin{figure}
	\begin{center}
		\includegraphics[width=0.49\textwidth]{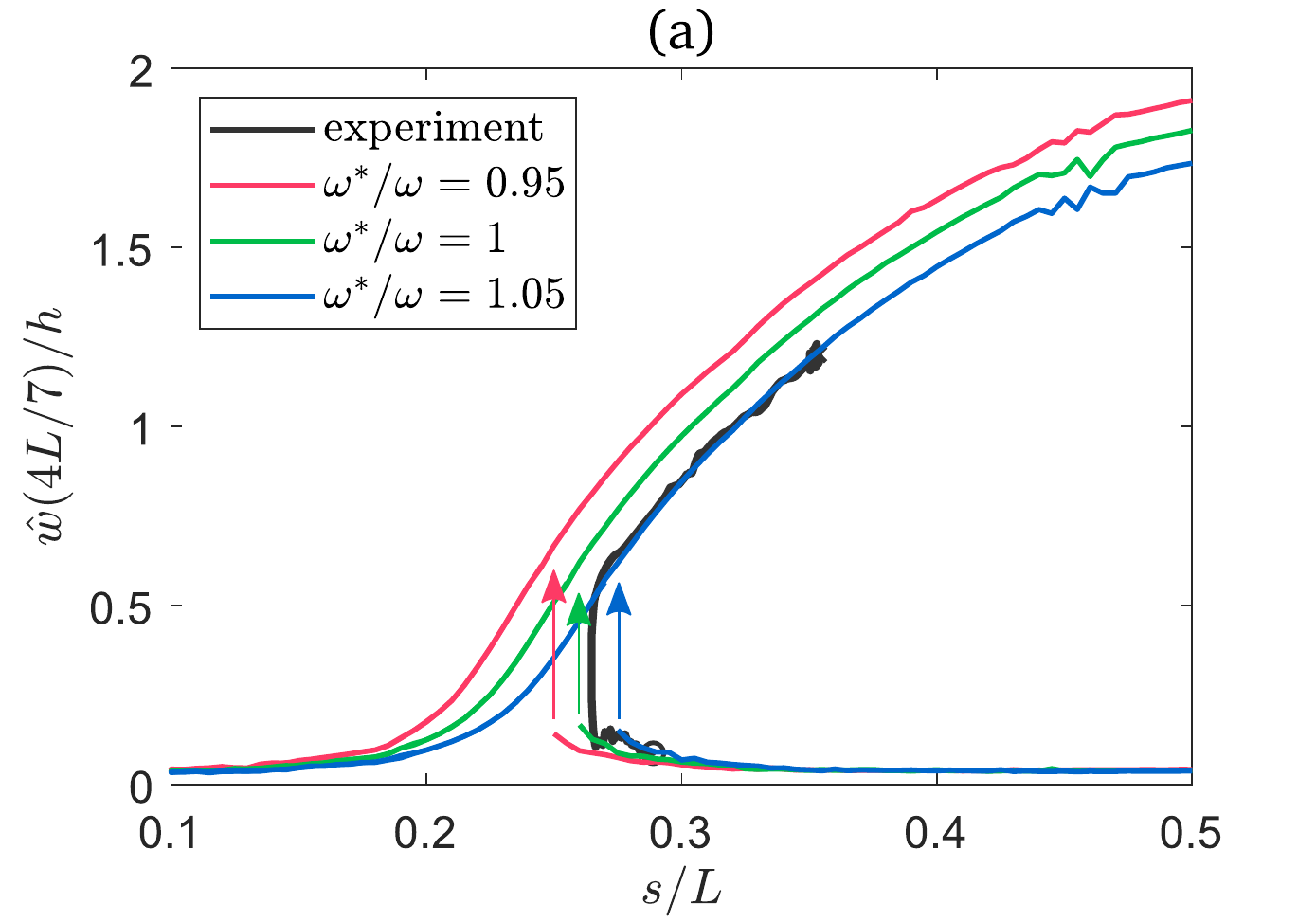}
		\includegraphics[width=0.49\textwidth]{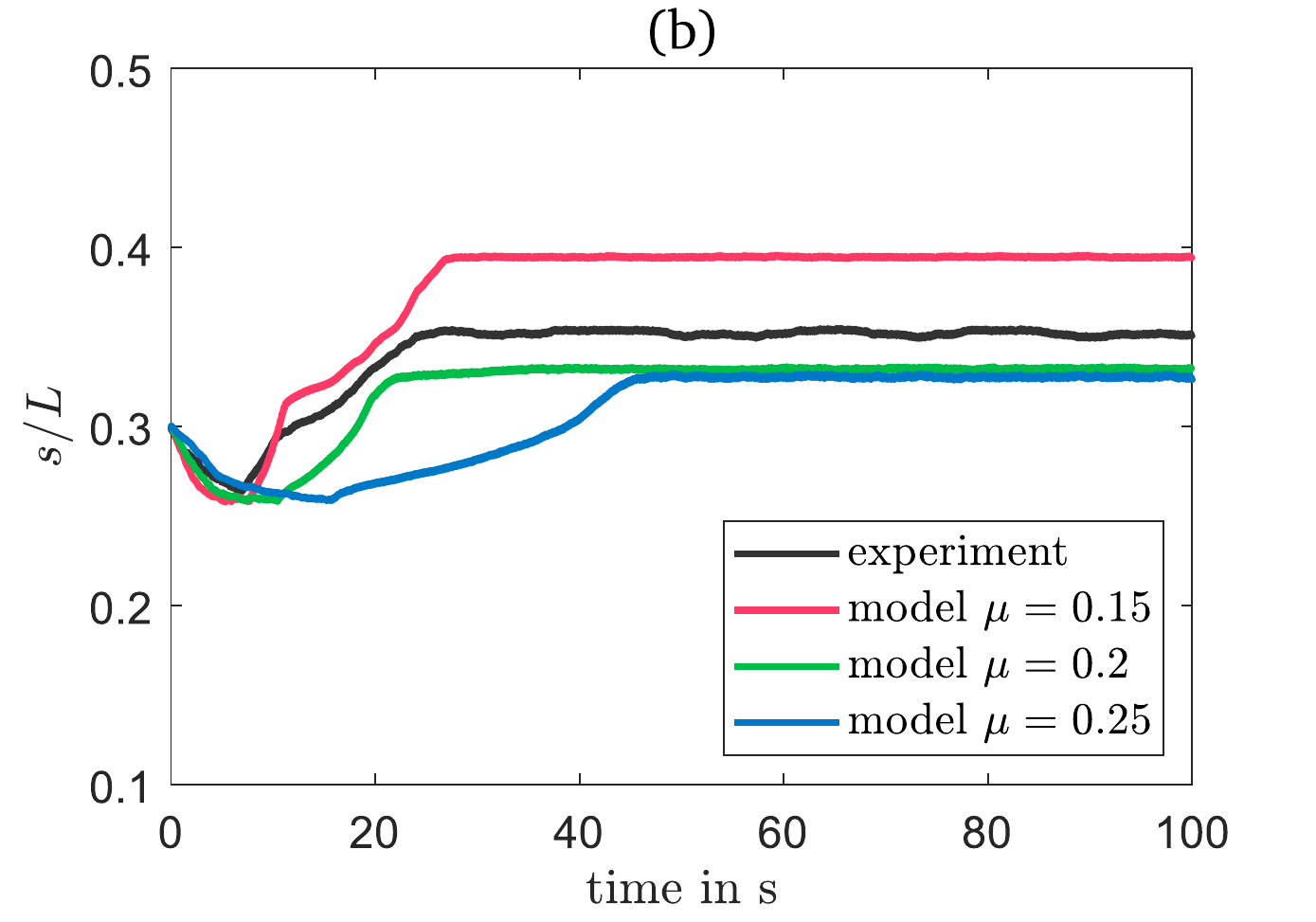}
		\includegraphics[width=0.49\textwidth]{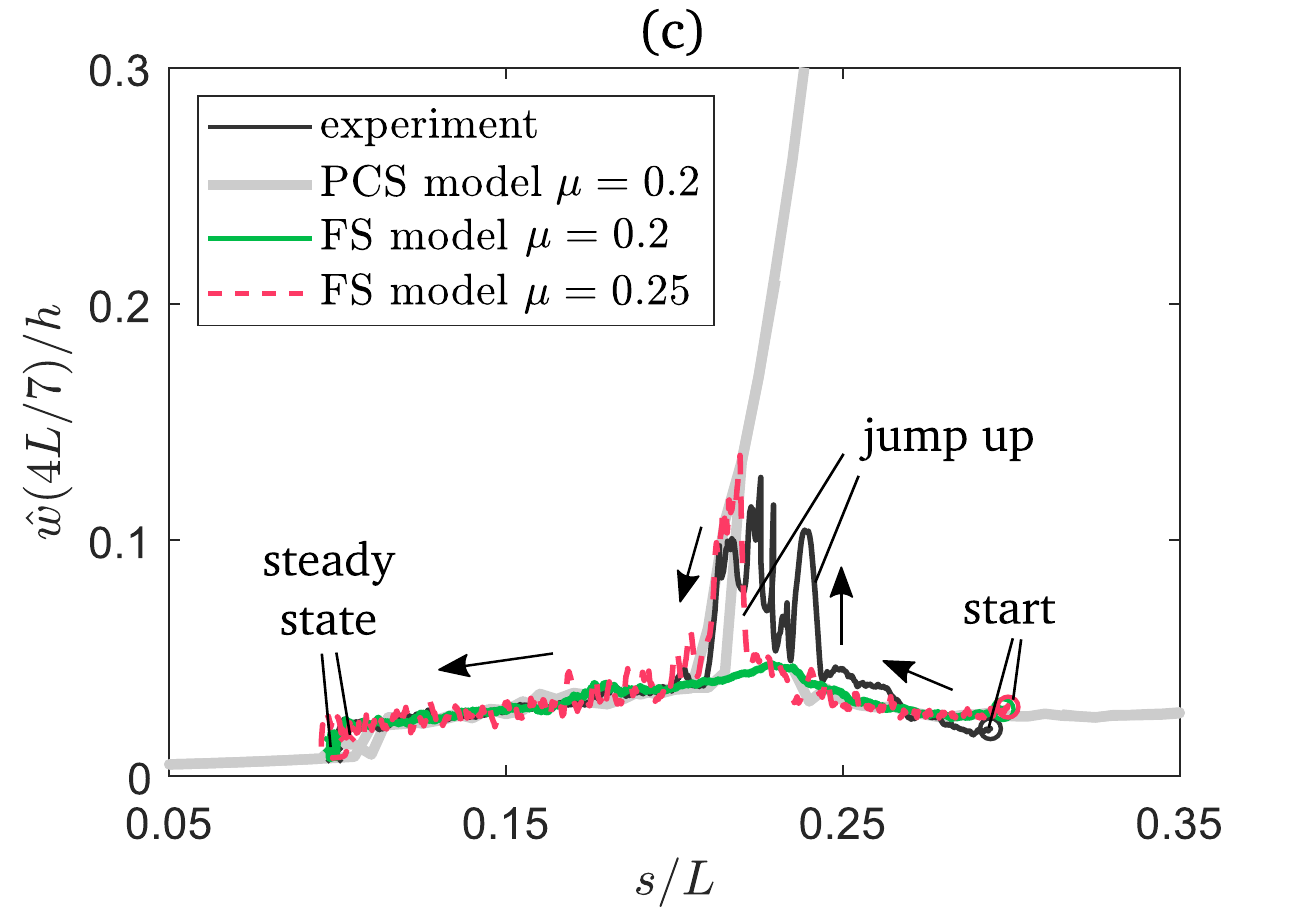}
		\caption{(a) Sensitivity of super-slow manifold with respect to the linear natural frequency: Experiment and PCS model at~$f_\mathrm{ex}=124\mathrm{Hz}$,~$\exa=14\mathrm{m/s}^2$ (b) slider position versus time under variation of friction coefficient: Signature move, same operating point \MOD{(c) super-slow manifold for the case of self-damping, reference and slightly increased coefficient of friction at~$f_\mathrm{ex}=124\mathrm{Hz}$,~$\exa=6\mathrm{m/s}^2$}
}
		\label{fig:sensitivity}
	\end{center}
\end{figure}
\\
The steady slider position is highly sensitive with respect to the friction coefficient as shown in~\fref{sensitivity}b.
As already indicated in~\sref{sim_prms}, the coefficient of friction is an uncertain parameter and it varies along the beam's length.
\MOD{Also the super-slow manifold in the case of self-damping depends on the friction coefficient. In the representative operating point (shown in \sref{Self-damping}), a slightly increased friction coefficient yields a bigger amplitude jump comparable to the one observed in the experiment, see \fref{sensitivity}c}. In conclusion, \MOD{the small quantitative deviations between experiment and model can be attributed} to the uncertainties inherent to jointed structures \cite{Brake.2018b}.
\subsection{Operating \MOD{regimes}}\label{sec:op_map}
Finally, \MOD{the operating map spanned by the excitation level and excitation frequency in \fref{op_map} is analyzed. The map} illustrates the type of behavior reached for any excitation parameter combination.
The discrete values of the excitation level are $\exa\in\{4,6,8,\dots,16\}\mrm{m/s}^2$, and of the excitation frequency, $f_\mrm{ex}\in\{100,102,104,\dots,130\}\mrm{Hz}$.
Besides ranges of the already discussed four types of behavior (signature move, trivial adaption, relaxation oscillation, self-damping), there is a small range of \emph{inactivity} where the slider does not move at all.
This range appears for relatively low excitation frequency.
Apparently, the beam-slider system is too far from resonance in this range, so that the vibration level is not sufficient to activate any locomotion of the slider.
Comparing experiment and simulation, this range appears at high and low excitation amplitude, respectively.
Simulations at lower excitation frequency (not captured in~\fref{op_map}) yield the state of inactivity also for large excitation level.
The rest of the map is filled by \MOD{well} separated ranges corresponding to the four described types of behavior.
From top left to bottom right, these ranges are arranged in the order adaption (trivial)--adaption (signature move)--relaxation oscillation--self-damping.
Experiment and model also agree with regard to this arrangement. \MOD{Considering the regime of relaxation oscillation in the model, the grid spanned by the operating conditions needed to be refined in order to show continuity.}
In the model, the range of the signature move is larger than in the experiment.
The shift of the border between trivial adaption and signature move is explained by the slightly shifted (potential) jumping point.
\MOD{It is worth emphasizing, that this border is also depending on the slider's start position. More precise, it is crucial wether one starts left or right of the potential jumping point, as illustrated for one operating point in~\fref{var_start_pos}.}
Also the borders between the other operating ranges are somewhat shifted, which \MOD{is attributed} to the uncertainties (in particular natural frequency, friction coefficient) and other imperfections (for instance imperfect excitation level control) inherent to the experiment, as already mentioned.
\begin{figure}
	\begin{center}
		\includegraphics[width=0.49\textwidth]{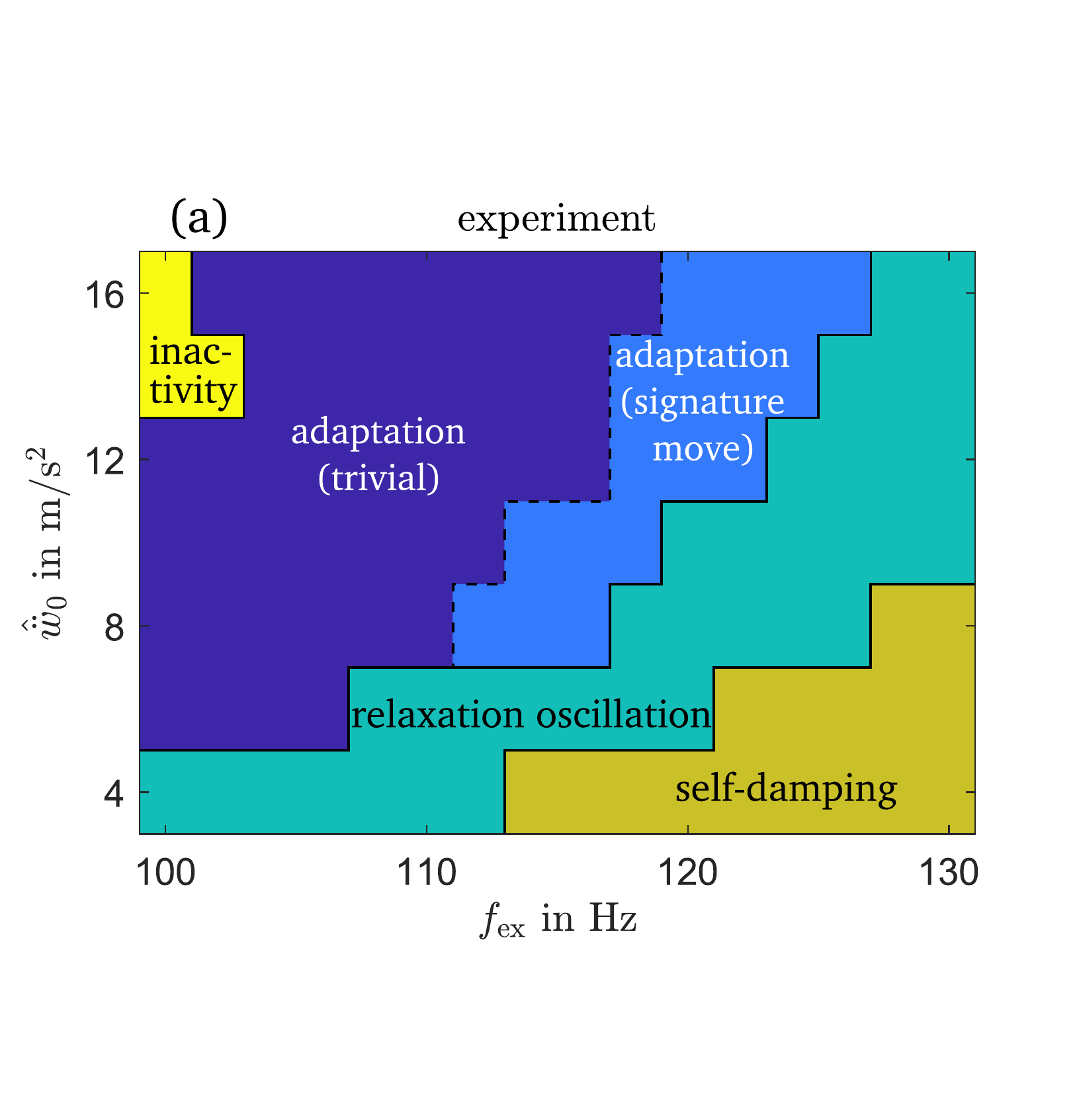}
		\includegraphics[width=0.49\textwidth]{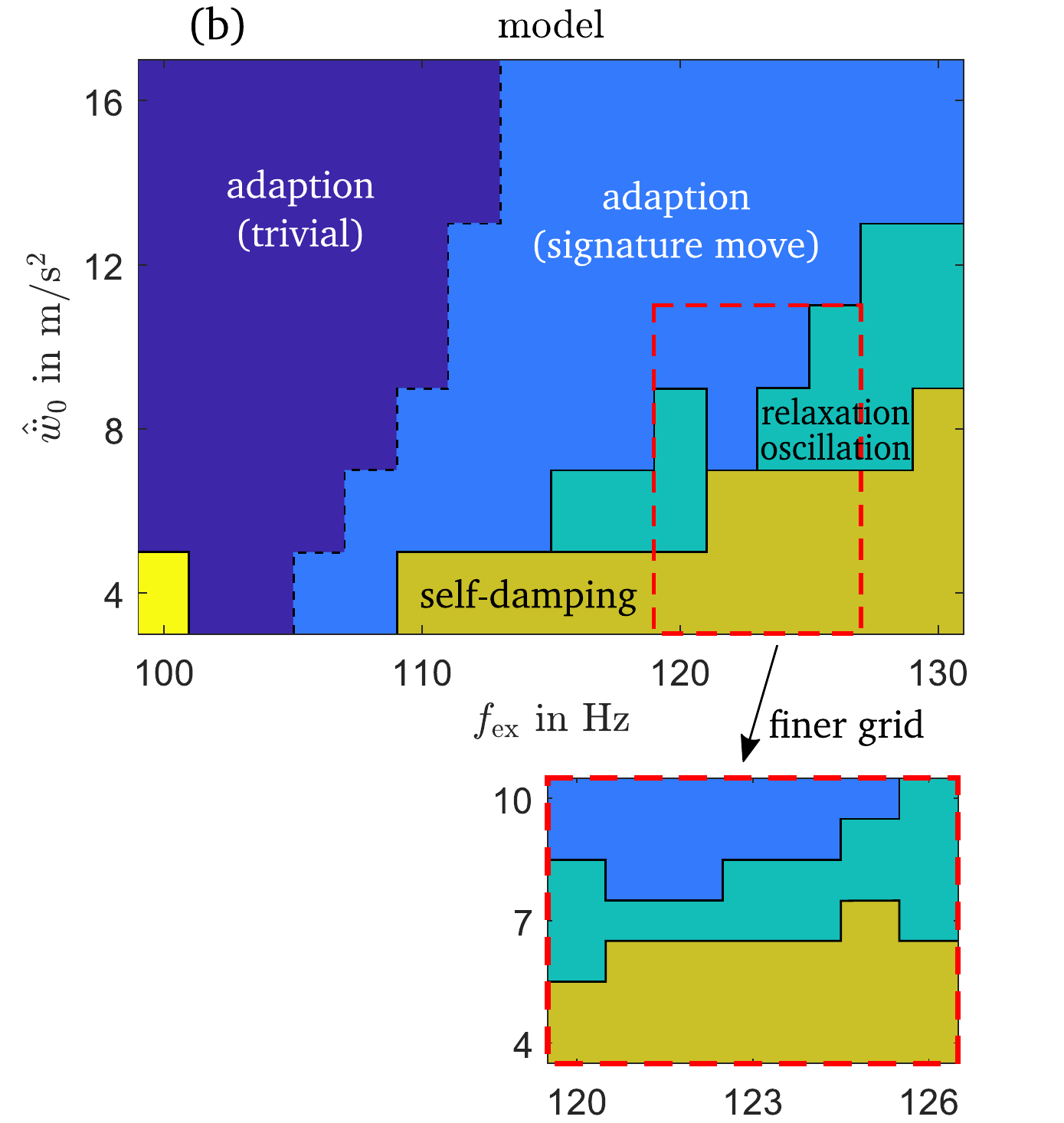}
		\caption{Operating map illustrating the ranges of different types of behavior (a) experiment (b) simulation
}
		\label{fig:op_map}
	\end{center}
\end{figure}
\begin{figure}
	\begin{center}
		\includegraphics[width=0.49\textwidth]{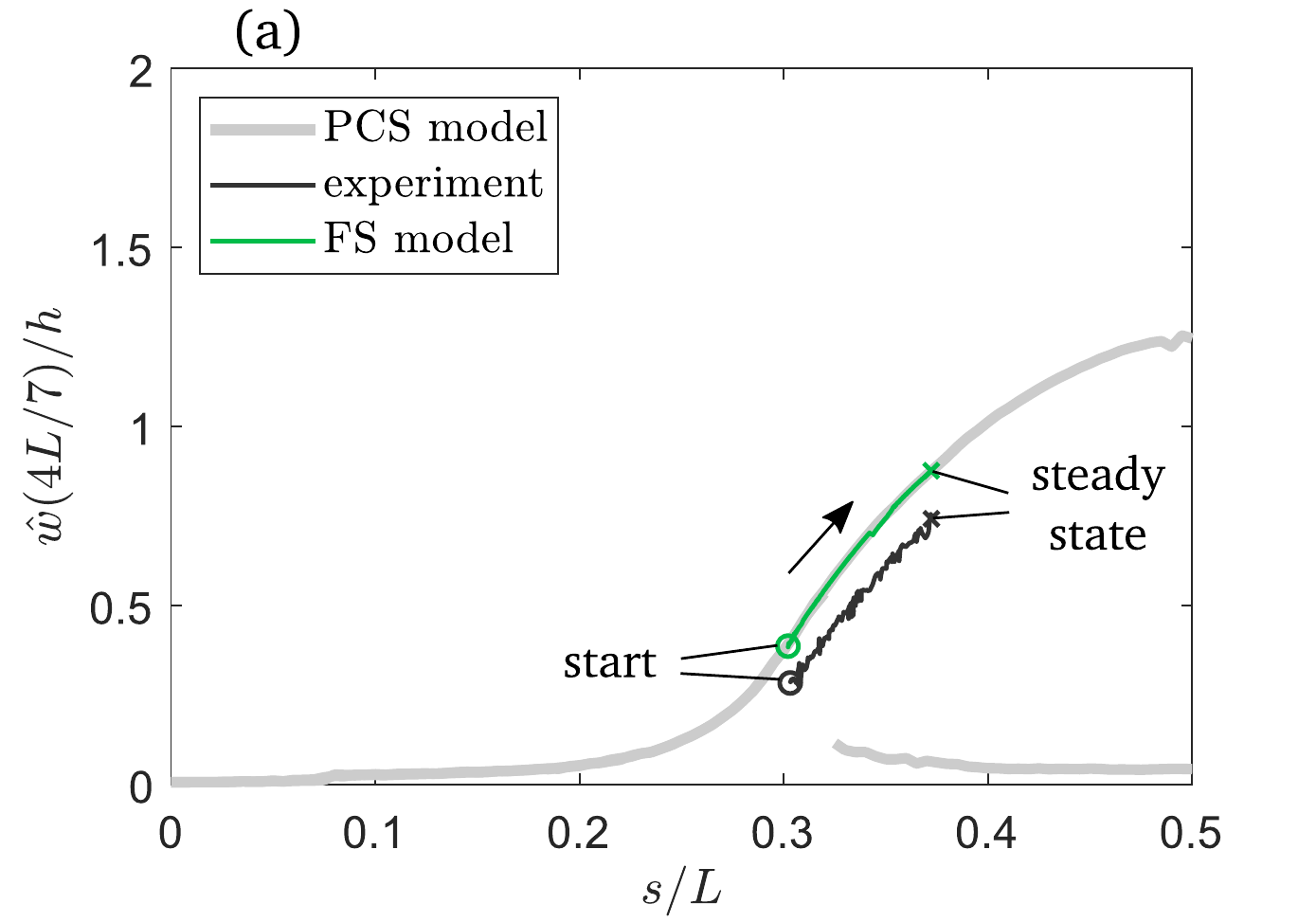}
		\includegraphics[width=0.49\textwidth]{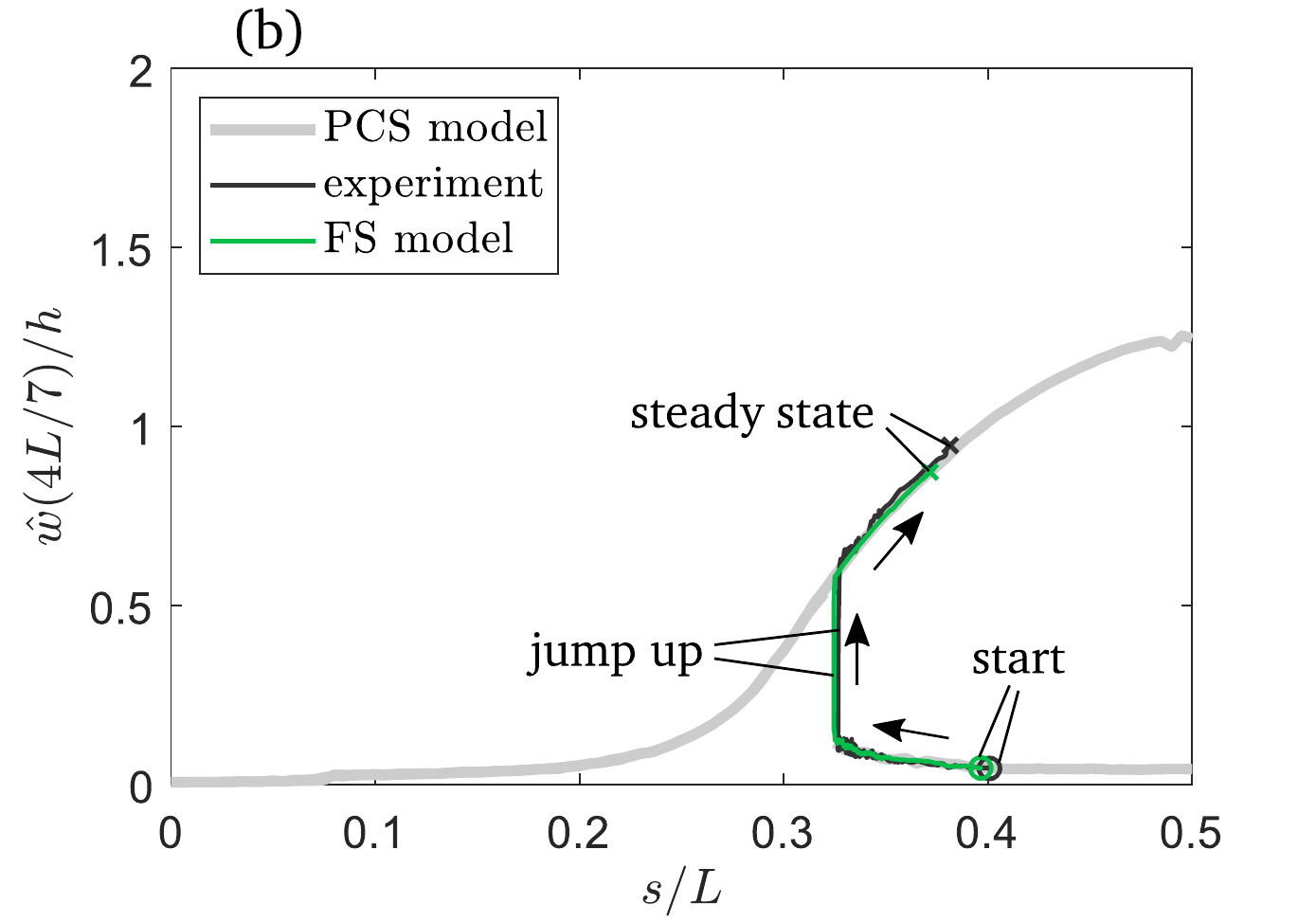}
		\caption{\MOD{Successful adaption under variation of slider's starting position at~$f_\mathrm{ex}=106\mathrm{Hz}$,~$\exa=10\mathrm{m/s}^2$ (a) trivial adaption starting from~$s/L=0.3$ (b) signature move starting from~$s/L=0.4$}}
		\label{fig:var_start_pos}
	\end{center}
\end{figure}
\section{Conclusions\label{sec:conclusions}}
Subject of the current work is the qualitative and quantitative validation of a model of a self-adaptive beam-slider system.
\MOD{A new prototype of the system with continuously adjustable clearance and concave inner slider geometry was designed in order} to facilitate robust locomotion.
All technical drawings are provided so that other research groups can easily reproduce the tests \cite{Muller.2022b}.
The new prototype yields successful adaption in a broad range of operating conditions.
In adjacent operating ranges, a relaxation oscillation and self-damping behavior are observed.
To accurately describe the dynamics of the geometrically nonlinear beam, \MOD{the model was refined} by considering the finite stiffness of the clamping.
The properties of the clamping stiffness and the modal damping were updated using the results of an Experimental Nonlinear Modal Analysis of the beam without slider.
Numerical simulation results of the beam-slider system, accounting for the clearance and the frictional contact interactions between the bodies, are in excellent agreement with the experiments.
Qualitatively, the model captures all characteristic types of behavior observed in the experiments (trivial adaption, signature move, relaxation oscillation, self-damping, inactivity) and all involved features. More precisely:
\begin{itemize}
\item[(1)] Movement of the slider away from/towards the beam's center at low/high amplitude
\item[(2)] Strongly modulated response at low amplitude
\item[(3)] Super-slow movement of slider compared to beam's vibration
\item[(4)] Jumps to higher/lower vibration level
\item[(5)] Maintenance of slider position and vibration amplitude at steady state after adaption
\item[(6)] Transition to harmonic vibration at very low amplitude
\end{itemize}
Quantitatively, experiment and model agree excellently in terms of:
\begin{itemize}
\item[(1)] Speed of the slider's horizontal movement
\item[(2)] Amplitude levels during corresponding phases
\item[(3)] Final slider position in case of transition to harmonic vibration
\end{itemize}
Minor deviations of the steady slider position and the super-slow manifold (vibration amplitude vs. slider position) are observed.
These are explained by the system's sensitivity to inevitable uncertainties, in particular with regard to the friction coefficient and the linear natural frequency.
\MOD{In conclusion}, the intended validation of the simulation model is entirely successful.
This provides a proper basis for further theoretical studies to gain deep understanding of the intriguing dynamics of the considered system and the model-based optimization for technical applications such as energy harvesting.

\end{document}